\documentclass[nofootinbib,aps,twocolumn,superscriptaddress,showpacs,english]{revtex4-1}

\usepackage[T1]{fontenc}
\usepackage[utf8]{inputenc}

\usepackage[english]{babel}
\usepackage{amsmath}
\usepackage{amssymb}
\usepackage{wasysym}
\usepackage{graphicx}
\usepackage{xcolor}
\usepackage{siunitx}
\usepackage{bm} 
\usepackage{cancel}
\usepackage{mdframed}

\usepackage{braket}

\usepackage[linktocpage=true,
  colorlinks=true, 
  pdfborder={0 0 0},
  linkcolor=blue,
  citecolor=red,
  filecolor=yellow,
  urlcolor=blue,
  bookmarks,
  pdfauthor={},
]{hyperref}

\usepackage{xcolor}   
\usepackage{pst-node} 

\newcommand{\GrazTh}{Institute of Theoretical and Computational Physics, Graz University of Technology, 8010 Graz, Austria}
\newcommand{\GrazEx}{Institute of Experimental Physics, Graz University of Technology, 8010 Graz, Austria}
\newcommand{\contributedequaly}{M.R. and P.H. contributed equally to this work.}

\newcommand{\eq}[1]{Eq. (\ref{#1})}

\newcommand{\BK}{{\cal I}}

\newcommand{\oBK}{\pi,{\cal I}}
\newcommand{\CS}{\boldsymbol \vert}
\newcommand{\oxi}{\overline \xi}
\newcommand{\xxx}[1]{{\color{blue}\cancel{#1}}}
\newcommand{\new}[1]{{{#1}}}
\newcommand{\uu}{1\hspace{-4pt}1}
\newcommand{\Ro}{^{(\rho)}}
\newcommand{\RR}{}

\begin{document}

\title[Analysis of femtosecond pump-probe PEPICO measurements]{Analysis of femtosecond pump-probe photoelectron-photoion coincidence measurements applying Bayesian probability theory}

\renewcommand{\thefootnote}{\fnsymbol{footnote}}

\author{M. Rumetshofer}  
\email{m.rumetshofer@tugraz.at} 
\affiliation{\GrazTh} 
\author{P. Heim} 
\thanks{\contributedequaly} 
\affiliation{\GrazEx} 
\author{B. Thaler}    
\affiliation{\GrazEx}
\author{W. E. Ernst}          
\affiliation{\GrazEx}
\author{M. Koch}              
\affiliation{\GrazEx}
\author{W. von der Linden}    
\affiliation{\GrazTh}

\date{\today}

\begin{abstract}
Ultrafast dynamical processes in photoexcited molecules can be observed with pump-probe measurements, in which information about the dynamics is obtained from the transient signal associated with the excited state. Background signals provoked by pump and/or probe pulses alone often obscure these excited state signals. Simple subtraction of pump-only and/or probe-only measurements from the pump-probe measurement, as commonly applied, results in a degradation of the signal-to-noise ratio and, in the case of coincidence detection, the danger of overrated background subtraction. 
Coincidence measurements additionally suffer from false coincidences, requiring long data-acquisition times to keep erroneous signals at an acceptable level. 
Here we present a probabilistic approach based on Bayesian probability theory that overcomes these problems. For a pump-probe experiment with photoelectron-photoion coincidence detection, we reconstruct the interesting excited-state spectrum from pump-probe and pump-only measurements. This approach allows us to treat background and false coincidences consistently and on the same footing. 
We demonstrate that the Bayesian formalism has the following advantages over simple signal subtraction: 
(i) the signal-to-noise ratio is significantly increased, 
(ii) the pump-only contribution is not overestimated, 
(iii) false coincidences are excluded,
(iv) prior knowledge, such as positivity, is consistently incorporated, 
(v)  confidence intervals are provided for the reconstructed spectrum, and 
(vi) it is applicable to any experimental situation and noise statistics. 
Most importantly, by accounting for false coincidences, the Bayesian approach allows us to run experiments at higher ionization rates, resulting in a significant reduction of data acquisition times. 
The probabilistic approach is thoroughly scrutinized by challenging mock data. The application to pump-probe coincidence measurements on acetone molecules enables quantitative interpretations about the molecular decay dynamics and fragmentation behavior.
All results underline the superiority of a consistent probabilistic approach over ad-hoc estimations.
The software implementation of the Bayesian formalism presented in this paper is provided at \url{https://github.com/fslab-tugraz/PEPICOBayes/}.
\end{abstract}

\maketitle

\section{Introduction}

Coincidence measurements are a widely used and powerful experimental technique in physics and chemistry. For example, in photoionization studies of gas phase molecules or clusters, photoelectron-photoion coincidence (PEPICO) detection, provides essential insights into the ionization process, which cannot be achieved by sole detection of ions or electrons~\new{\cite{Boguslavskiy2012,Sandor2014,Koch2017,Arion2015,Continetti2001}}. Introduced in the 1960s~\cite{Brehm1967}, coincidence methods have rapidly developed and are nowadays also applied in time-resolved investigations of ultrafast dynamics in molecules or clusters. In these dynamical studies PEPICO detection has proven to be essential to learn about the underlying processes if competing intramolecular relaxation pathways are active~\cite{Maierhofer2016,Koch2017,Couch2017,Wilkinson2014}, or if different species are present~\cite{Hertel2006}.

\begin{figure}[h!]
    \centering
    \includegraphics[width=1.0\columnwidth,angle=0]{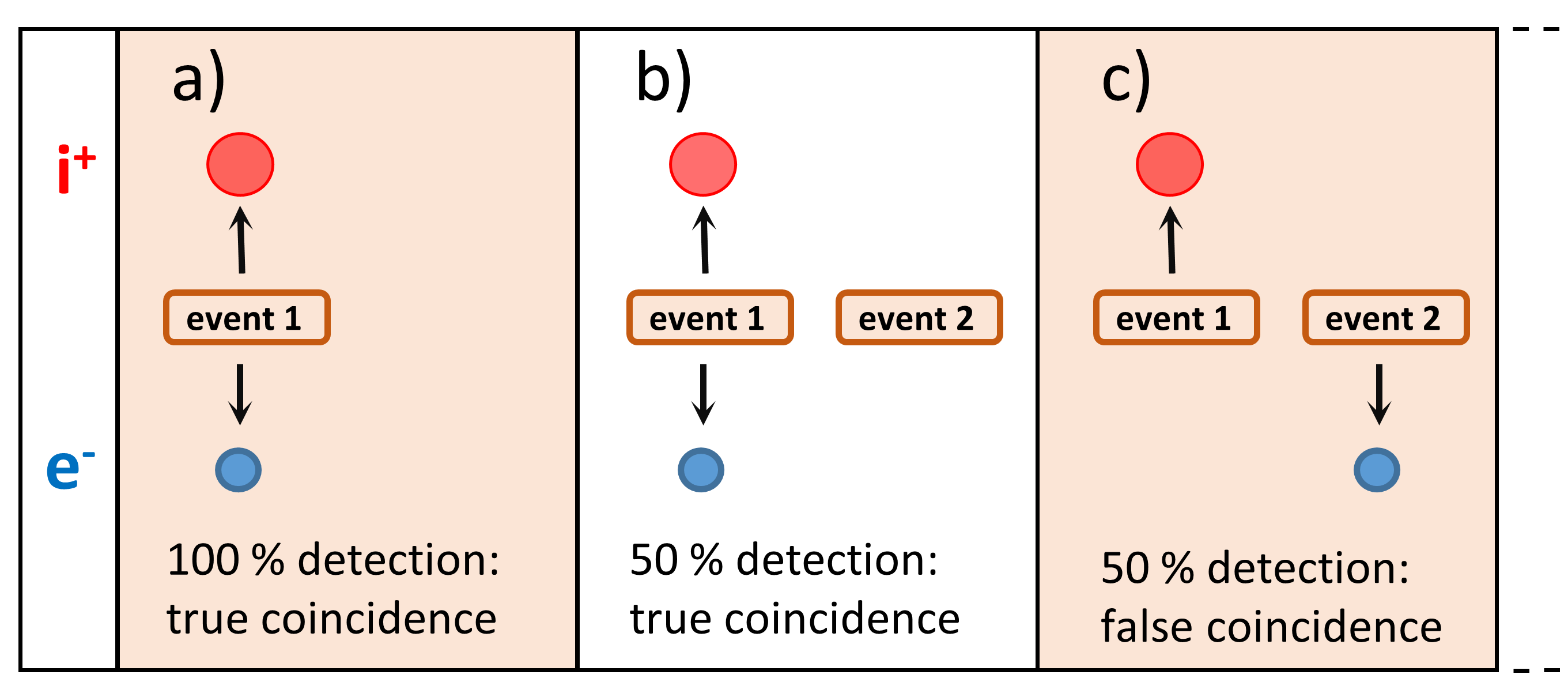}
    \caption{Three possible outcomes if one or two molecules get ionized in a coincidence measurement.
    In the ideal case (a) one event is generated and the created electron-ion pair is detected.
    If the detection probability is \new{less than one} and one electron and one ion are detected it can either be a true coincidence, if both stem from the same molecule, example (b), or a false coincidence, if they originally belonged to different molecules, example (c).}
    \label{fig:comsubbayes1}
\end{figure}

While the success of PEPICO detection is based on the unambiguous recording of pairs of energy-resolved electrons and the corresponding mass-resolved cations, the correct pairwise assignment (true coincidence) may be affected by certain experimental conditions: If a laser pulse triggers a number of simultaneous ionization events arising from different neutral molecules, and if the detection probability is imperfect, the assignment of correlated electron-cation pairs suffers and gives rise to so-called false coincidences~\cite{Stert_1999}. 
\new{In principle there are also false coincidences due to detector noise or ionization events not caused by the laser pulse, but these  { are sufficiently low to be neglected in the presented experiment and}  are { therefore } not covered in this paper.}
The issue of false coincidences is exemplified in Fig. \ref{fig:comsubbayes1}, which shows that even restricting the recording to  single ionization events, can yield a wrong correlated pair assignment  due to low detector sensitivity.
\new{{Momentum imaging techniques}, such as cold target recoil ion momentum spectroscopy (COLTRIMS) \cite{Doerner2000,Ullrich2003}, are in principle able to account for false coincidences. Based on exact spatial detection of all fragments and the reconstruction of their initial momentum vectors after ionization and fragmentation, {these methods} allow one to filter for ionization events that fulfill momentum conservation, i.e., originate from one molecule. 
{However, time-of-flight detection, which is applied in the presented experiment to detect photoelectrons with high energy resolution, does not allow for identification of false coincidences based on experimental observables.}}

Misinterpretation can be avoided by the method of covariance mapping, which is based on the calculation of the covariance for the photoelectron and mass spectra measured with each laser shot~\cite{Frasinski1989,Mikosch2013b,Mikosch2013c}. However, covariance mapping does not guarantee that the reconstructed spectrum is positive and it is restricted to Poisson processes and leads otherwise to systematic deviations~\cite{Mikosch2013b,Mikosch2013c}. Further limitations are outlined in Ref. \cite{Frasinski1989}. 
Of course, keeping the average number of simultaneous ionization events far below one to avoid false coincidences~\cite{Stert_1999} also serves the purpose but requires long data acquisition times for sufficient signal-to-noise ratios. This restriction can be circumvented by the Bayesian approach, which will be presented in this paper.  Bayesian probability theory is the consistent approach to reconstruct spectra from any noisy experimental data~\cite{jaynes_probability_2003} and the reconstructed spectra are never negative. Moreover, the probabilistic approach provides confidence intervals, which are crucial to assess the reliability of structures in the reconstructed spectrum.
Also the issue of false coincidences can be dealt with consistently.
The Bayesian approach can also overcome
another problem that arises in pump-probe PEPICO, which is related to the fact that two large signals with significant statistical fluctuations have to be subtracted. The situation is as follows: Time-resolved studies are carried out as pump-probe experiments~\cite{Hertel2006,Stolow2004}, where the photoexcitation by a pump pulse triggers dynamical processes in the electronic and nuclear structure of the molecule. A time-delayed probe pulse photoionizes the molecule and the transient change of photoelectron and -ion signals associated with the excited states provide insight into the underlying processes. Unfortunately, also pump and/or probe pulses on their own, referred to as pump-only and probe-only pulses, can lead to photoionization, resulting in a background signal that is superimposed on the excited state signal. If possible, the laser intensity of the pump and the probe pulses is reduced to minimize this background signal. However, often the pump-only and/or the probe-only signal significantly contribute to pump-probe measurements, particularly if multiphoton transitions are applied for pump excitation or probe ionization, or if high photon energies are used for probing~\cite{Koch2015,Koch2014c,Nugent-Glandorf2001}. To obtain the true excited-state transients, the pump-only and/or the probe-only signals are separately measured and usually subtracted from the pump-probe signals, resulting in increased noise in the obtained spectra. The increase in noise is particularly severe if the pump-probe signal cannot be spectrally separated from the pump-only and probe-only signals, that is, if the respective spectra overlap.

Additionally, it has to be considered that the pump-only, the probe-only, and the pump-probe measurements have different rates of ionization events (see Sec. \ref{sec:notation}). Simple subtraction of the signals leads to errors because the coincidence signals depend on the ionization rates with the consequence that pump-only and probe-only measurements are different from the pump and probe contribution in the pump-probe measurement.
Moreover, we note that also population depletion effects can change the rate of ionization events from certain states, in particular if the probe pulses lead to ionization of the ground state. For instance, photoexcitation by the pump pulse reduces the ground state population of a molecule with the consequence that the probe pulse ionization rate corresponding to the ground state is reduced and thus lower than the ground state signal of the probe-only measurement.
In this work, we apply Bayesian probability theory to infer the underlying time-dependent excited-state dynamics in the presence of a strong pump-only background and a negligible probe-only signal, that is from pump-probe and pump-only measurements.
Depletion effects associated with excited states, similar to the mechanisms described above, occur only if pump and probe pulses overlap in time. 
Since we are interested in the dynamics of the photoexcited states after the pump excitation is completed, we focus on measurements with temporarily separated pulses and therefore neglect depletion effects in the current approach. 

 The application of the Bayesian formalism for background subtraction was presented for astrophysical applications~\cite{Gregory2005,Loredo1992} and for photo-induced x-ray emission spectroscopy (PIXE)~\new{\cite{Gertner_PIXE_1989,von_der_linden_signal_1999,von_der_linden_how_1996,prozesky_use_1997,padayachee_bayesian_1999,fischer_background_2000}}. Compared to conventional subtraction of the pump-only spectrum from the pump-probe spectrum, the Bayesian approach provides several important advantages:
(i) It results in a significant increase of the signal-to-noise ratio.
(ii) It does not overestimate the pump-only contribution and does never lead to negative spectra because the relative weight of the pump-only contribution is self-consistently determined. 
\new{(As explained below, the experimental conditions are such that the pump pulse excites ground
state molecules and the probe pulse ionizes exclusively excited states, with the consequence that a pump-probe
measurement yields always more or equal ionization events compared to the pump-only measurement. { But more ionization events in the pump-probe measurement can lead to fewer single coincidence events, and therefore to overestimation of the pump-only contribution and to negative difference spectra.})}
(iii) Spectral signatures based on false coincidences are eliminated, allowing for higher signal rates.
(iv) It includes consistently all prior knowledge, such as positivity, and
(v) a confidence interval is obtained for the estimated spectrum.
(vi) It is applicable to any experimental situation and noise statistics.
We provide our software, including introductory examples, at
\url{https://github.com/fslab-tugraz/PEPICOBayes/}.

\section{Experiment and Assumptions\label{sec:two}}
\label{sec:exp}

\begin{figure}[h]
\centering
\includegraphics[width=1.0 \columnwidth]{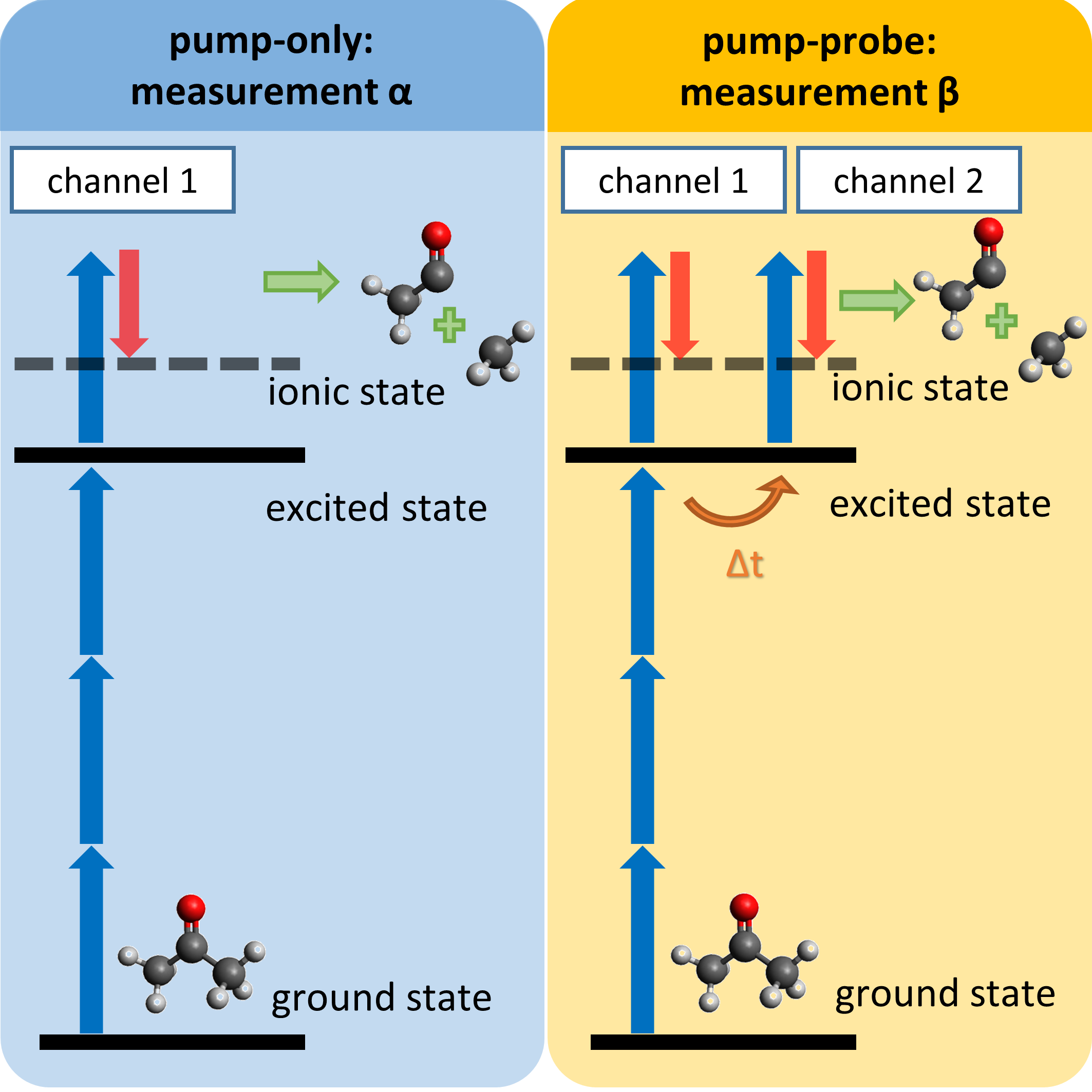}
\caption{Pump-probe ionization scheme to investigate excited-state dynamics in molecules. Left: Ground-state molecules are ionized by pump pulses alone (measurement $\alpha$). Right:  The combination of pump and time-delayed ($\Delta t$) probe pulses (measurement $\beta$) results in pump-only (channel 1) and excited-state ionization events (channel 2). Red arrows indicate the electron kinetic energy and potential fragmentation is also depicted.}
\label{fig:scheme}
\end{figure}

The goal of the time-resolved PEPICO experiments is to determine the time-dependent excited-state population of a particular molecule. 
In a femtosecond pump-probe measurement a fraction of ground-state molecules is excited by a pump pulse and subsequently ionized by a time-delayed probe pulse. 
The time-resolved  distribution of the electron kinetic  energy provides valuable information about dynamical properties of the electronic structure. The simultaneous detection of the ion-mass allows to assign
the electronic features to a particular molecule. 
If multiphoton transitions are applied for the excitation or ionization transition, the pump or probe laser pulse, respectively, causes a strong background signal for the pump-probe measurement.
In this work, we consider the case of a strong pump-only background. Due to the low laser intensity of the probe pulse, ground-state molecules are not ionized by the probe pulse alone and there is no probe-only background.
This situation is shown in Fig. \ref{fig:scheme} for a three-photon excitation to high-lying molecular states. The excited state lies energetically close to the ionization continuum, resulting in a certain probability for four-photon ionization -- the background signal -- in addition to the three-photon excitation (measurement $\alpha$ and channel 1 in Fig. \ref{fig:scheme}). 

In a separate pump-probe measurement ($\beta$) the pump process is the same as in the pump-only case. It generates  excited states which in turn are ionized by a time-delayed probe pulse. Consequently, the measured pump-probe spectrum (measurement $\beta$) consists of both, pump-only ionization events (channel 1) and pump-probe events (channel 2). 
 
Photoelectrons and -ions are both detected with high efficiency by a time-of-flight spectrometer, where the electron kinetic energy and the ion mass are measured~\cite{Maierhofer2016,Koch2017}.
In dependence on the ionization path, cations produced in both channels can be stable and detected as parent ions or undergo fragmentation into neutral and ionic fragments.
Coincidence detection of electrons and ions allows us to obtain separate electron spectra for each ion fragment. 
{ The excited electronic state of the molecule at the moment of probe ionization is identified by the measured electron kinetic energy, in combination with the energy of the ionizing photon and knowledge of the vertical ionization energy of excited state. 
In addition to the information of species and electronic state that is ionized, the related ion mass of the PEPICO spectrum provides insight into the fragmentation behavior.}
\new{For example, the assignment of the photoelectron kinetic energy to an excited electronic
state of the unfragmented molecule and coincidence detection of an ion fragment shows that the molecule was
intact at the moment of ionization and fragmentation must have occurred in the ionic state. This channel plays
an important role in the results on acetone, as presented below.}
PEPICO detection thus allows us to disentangle different relaxation and ionization pathways in photoexcited molecules~\cite{Maierhofer2016,Koch2017,Wilkinson2014}.
 
For coincidence detection, only events are considered, in which one electron-ion pair is detected,  assuming that both result from the same molecule. If molecules are ionized within one laser pulse, the possible options are (compare with Fig. \ref{fig:comsubbayes1}): No single electron-ion pair is detected, in this case the event is rejected, or one electron and one ion are detected, which can originate [Figs. \ref{fig:comsubbayes1}(a) and \ref{fig:comsubbayes1}(b)] from the same molecule (true coincidence) or [Fig. \ref{fig:comsubbayes1}(c)] from different molecules (false coincidence). Since the pump-probe measurement ($\beta$) has a higher ionization rate compared to the pump-only measurement ($\alpha$), the number of single coincidences in channel 1 differs from that in the pump-only measurement.
\new{In other words, in a pump-probe measurement the ions and/or electrons originating
from ionization of photoexcited molecules (channel 2) are detected with a certain probability with electron-ion
pairs originating from the ground state (channel 1), in which case the event is discarded. As a result, the
number of registered channel 1 events of a pump-probe measurement is always lower than that of a pump-only
measurement.}
Consequently, simply subtracting  the pump-only counts from  the pump-probe counts would lead to wrong results.

Moreover, the populations in the excited state can decay to energetically lower states by fast and efficient nonadiabatic processes~\cite{Maierhofer2016,Koch2017,Koch2017a}. The channel 2 signal in a pump-probe measurement can therefore become significantly smaller than the channel 1 background, in particular for long delay times. Especially in this situation, simple subtraction of the pump-only from pump-probe counts results in a very poor signal-to-noise ratio.

Before applying the Bayesian formalism, which is presented in Sec. \ref{sec:bayes}, to real experimental data, it will be tested
by means of some challenging mock data in Sec. \ref{sec:mock}.
Then, the investigation of photoinduced relaxation dynamics of acetone molecules is presented in Sec. \ref{sec:application} to demonstrate the application of the Bayesian formalism.

For the measurements we use a femtosecond pump-probe setup, which has been described in detail previously~\cite{Maierhofer2016,Koch2017a}. 
Acetone molecules are excited by a three-photon transition to high-lying Rydberg states (6$p$, 6$d$, 7$s$) at about 9.30 eV~\cite{Koch2017a}. 
Pump and probe pulses are obtained from a commercial Ti:sapphire laser system (Coherent Vitara oscillator and Legend Elite Duo amplifier) and frequency doubled in BBO crystals to obtain 3.1 eV photon energy (395-nm center wavelength). 
\new{The pump-probe cross correlation was ($84 \pm 1$) fs full width at half-maximum.}
Acetone molecules are introduced into the vacuum chamber and ionized in the extraction region of a time-of-flight spectrometer, which is operated in a magnetic bottle configuration for electron detection and coincidence detection of ions is achieved by a pulsed electric field. The electron and ion flight times are first analyzed by a coincidence algorithm producing the data sets $D_1$ and $D_2$ (see below), from which the excited-state spectrum is reconstructed by the Bayesian algorithm, based on pump-probe and pump-only measurements.

\section{Some preliminary considerations}
\label{sec:notation}

First, we want to briefly introduce Bayesian probability theory. We suggest
Refs. \new{\cite{jaynes_probability_2003,Gregory2005,Sivia2006,Ghosh_2006,Hoff_2009,Jackman_2009,linden_bayesian_2014}} for a more detailed introduction into Bayesian probability theory.
As has been aptly described by Jaynes \cite{jaynes_probability_2003}, probability theory forms the logic of science. 
\new{Bayesian probability theory can be seen as the generalization of  Boolean algebra. It is based on propositions, i.e.,  statements that are either true or false. For example, $M$ is short hand for the proposition that 
{\it the measured mass is $M$}, and
{\it one elementary event  during measurement $\rho$} states that  during one measurement of type $\rho$,
only one elementary event happens. 
As in Boolean algebra, proposition can be combined by the logical OR ($\vee$) and the logical AND ($\wedge$).}
The proposition $\BK$ stands for the so-called {\it background information } that includes all additional information, which uniquely defines the data-analysis problem. It includes  the relation between the desired spectra along with unknown parameters and the experimental data, as well as the statistics of the experiment and any sort of additional prior knowledge. \new{For more details see
Refs. \cite{jaynes_probability_2003,Gregory2005,Sivia2006}.}
For notational ease, conjunctions are denoted by commas,
e.g., $P(A\wedge B) \to P(A,B)$.
The quantity $P(A|B)$ stands
for the conditional probability that proposition $A$ is true,
provided $B$ is true. 
Generally, Bayesian probability theory can be fully derived by quantifying the principles of logical consistency \new{\cite{Cox_1946,Kolmogorov_1950,Knuth_Bayes_2012}}. This leads to two basic rules, which will be exploited intensively in this paper. The first one is the sum rule, 
\begin{equation}
P(A\lor  B|C) = P(A|C) + P(B|C)\;,\quad\text{if } A\wedge B = 0\;,
\end{equation}
and the second one the product rule,
\begin{equation}
P(A,B|C) = P(A|B,C) \; P(B|C)\;.
\end{equation}
Combining the two leads to the marginalization rule,
\begin{equation}
P(A|C) = \sum_{B_{i}} P(A|B_{i},C) \; P(B_{i}|C)\;,
\end{equation}
provided, the propositions $B_{i}$ are pairwise exclusive
$B_{i}\wedge B_{j} = 0 \;,\quad\forall i\ne j\;,$
and the union of all proposition is the true proposition
\new{$\lor_{i} B_{i}  = 1$}.
But the most important consequence  of the product rule is Bayes' theorem,
\begin{equation}
    P(H|D,\BK) = \frac{P(D|H,\BK)\;P(H|\BK)}{P(D|\BK)}\;,
\end{equation}
which constitutes the rule for learning from 
experimental data $D$. The proposition $H$ stands for unknown quantities, such as the energy-resolved spectrum $q(E)$ or unknown parameters, e.g., the intensity of a Poisson distribution $\lambda$. In this context,
$P(H|\BK)$ is called {\it prior probability} and represents the prior knowledge about the unknown quantities  $H$,
conditional on additional information, which might be present in the background information $\BK$, such as additional parameters or the positivity of the spectrum. 
The {\it likelihood} $P(D|H,\BK)$, representing the probability for the data $D$ given  $H$, includes all information about the measurement itself. \new{For example, when dealing with Poisson processes,
the unknown  parameter is the true mean $H\to \lambda$ and the measured quantity $D\to m$ are the counts, which have the probability distribution ${\cal P}(m|\lambda)$ defined in \eq{eq:Poisson}}. 
\new{Another ubiquitous example for a likelihood 
is obtained in the case of additive noise. In this case, the underlying true value of a
physical quantity, $x$ say, is distorted by some noise $\eta$,
resulting in the experimental data $D\to d =x+\eta$.
In many cases $\eta$ is  Gaussian  distributed and $p(d|x,\sigma,\BK)$ is a Gaussian in $d-x$ with variance $\sigma^{2}$.
Here $d$ actually stands for a continuous quantity and
$p(d|x,\sigma,\BK)$ is 
a probability density function (\new{PDF}). 
Throughout this paper we will use lower case $p(.)$
for \new{PDF}.
In the Bayesian frame, a continuous variable, $\hat x$ say, is treated by propositions as follows. Let  $D_{x}$ stand for the proposition: ``the variable $\hat x$ has a value between $x$ and $x+dx$''; i.e., $\hat x\in (x,x+dx]$. Then $P(D_{x})$ stands for the probability that $\hat x$ has a value in $(x,x+dx]$. This in turn is expressed by $P(D_{x}) = p(x) dx$ and defines the \new{PDF} $p(x)$. }
In view of the considerations above, 
the likelihood is also termed {\it forward probability}, because knowing $H$ allows us to determine the probability for  $D$.
For any experimental setting, for which the likelihood can be
specified,
Bayes' theorem allows us to solve the inverse problem and determines the probability for unknown quantities $H$.
The denominator in Bayes' theorem, also named {\it data evidence}, ensures the correct normalization, and can be evaluated via the marginalization rule. 
\new{It is common in Bayesian probability theory to use a short hand notation for propositions, 
e.g., in the case of a discrete variable $m$ the
proposition $\hat m = m$, which means {\it the  variable $\hat m$ has the value $m$}, is simply expressed as $m$.
Likewise, for a continuous variable, $x$ say,
the proposition $\hat x = x$, which means {\it the  variable $\hat x$ has the value $x$}, is simply expressed as $x$ in \new{PDF}s.} This allows a much more concise notation and misinterpretations can easily by avoided. In this notation, the marginalization rule can, e.g., have the form
\begin{equation}
 P(A|\BK) = \int P(A|q,\BK) \; p(q) dq \;. \label{eq:marg}
\end{equation}
For more details see Refs. \cite{jaynes_probability_2003,linden_bayesian_2014,Gregory2005} for a detailed introduction into Bayesian probability theory.
\new{Readers, particularly interested in the foundations of probability theory,
may want to look at the work of Kolmogorov 
or more recent developments by Skilling and Knuth \cite{Kolmogorov_1950,Knuth_Bayes_2012}.
The difference between Bayesian probability theory and the ``frequentist'' point of view, can be found in Refs. \cite{Jackman_2009,Hoff_2009,linden_bayesian_2014}.}

Now we consider the following standard setup consisting of two experiments on the same target: pump-only and pump-probe, denoted by $\alpha$ and $\beta$, respectively.
Each experiment consists of ${\cal N}_p$ measurements, a measurement of the $\alpha$ experiment consists of one laser pulse, while 
in the $\beta$ experiment the measurement comprises a pump pulse and a probe pulse.
We refer to one measurement of the $\alpha$ or $\beta$ experiment as $\alpha$ or $\beta$ measurement, respectively.
During one measurement, two types of elementary coincidence events are detected, either a molecule is ionized from its ground state (referred to as channel 1) or from its excited state (channel 2).  The latter is only possible in the pump-probe measurement ($\beta$).
We assume that the number $m$ of elementary events in a single laser pulse is Poisson distributed with some mean $\lambda$. In this paper we assume that the laser intensity
is constant during the entire experiment and hence, $\lambda$ is the same for all laser pulses.
Mikosch et al. \cite{Mikosch2013b,Mikosch2013c} proposed to describe the  fluctuations of the 
laser intensity  by a Gaussian probability density function (\new{PDF}) for the individual $\lambda$ values.
It is straight forward, but needs a bit more mathematics, to allow for such fluctuating laser intensities in the Bayesian analysis. 
We assume that in each experiment, characterized by a defined delay time between pump and probe pulse, $\lambda$ is independent of the occupation of the spectrum, which means that we neglect population depletion effects.
In one elementary event, the involved molecule can have mass $M_{\mu}$ and the emitted electron energy $E_{\nu}$.
For brevity we will refer to  this particular event as $(\mu\nu)$.
{ The ion masses and the electron energies are discretized, $\mu,\nu\in\mathbb{N}$, due to the finite resolution of the time-of-flight spectrometer.}
We will also use the symbol $\rho$, if we refer to the measurements/experiments $\alpha$ or $\beta$, the symbol $j$ 
for the channels $1$ or $2$, and $x$ for the combination of both sets, i.e. $x\in\{1,2,\alpha,\beta\}$.
Given an elementary event happens during measurement $\rho\in\{ \alpha , \beta\}$, the probability that it corresponds  to $(\mu\nu)$ is denoted by
\begin{widetext}
\begin{eqnarray}
  q^{(\rho)}_{\mu\nu} &= P(M=M_{\mu},E=E_{\nu}\CS \text{one elementary event during measurement } \rho,\BK)
\end{eqnarray}
Moreover, we introduce 
\begin{eqnarray}
 q^{(j)}_{\mu\nu} &= P(M=M_{\mu},E=E_{\nu}\CS \text{one elementary event in channel }j,\BK)
\end{eqnarray}
\end{widetext}
the probabilities for $(\mu\nu)$ when an elementary event happens in channel $j \in\{1, 2\}$.
All probabilities are properly normalized,
\new{\begin{equation}
\sum_{\mu\nu} q^{(x)}_{\mu\nu} = 1 ~~~~ \forall x\in\{ 1,2,\alpha,\beta \}\;.
\end{equation}}
\new{We define the mean number of elementary events in a single laser pulse of the channels and measurements as $\lambda_x$ for all $x\in\{ 1,2,\alpha,\beta \}$.}
In the pump-only measurement ($\alpha$), all molecules are in their respective ground state, therefore only channel 1 is allowed,
\begin{equation}\label{eq:q:alpha}
q^{(\alpha)}_{\mu\nu} = q^{(1)}_{\mu\nu}\;,
\end{equation}
and $\lambda_\alpha = \lambda_1$.
\new{If an elementary event $(\mu\nu)$ happens in the pump-probe measurement ($\beta$), the event can belong to channel 1
or  2, with the respective probabilities $p^{}_{1}$ and $p^{}_{2}=1-p^{}_{1}$.}
Hence, we get
\begin{equation}\label{eq:q:beta}
q^{(\beta)}_{\mu\nu} 
=
p^{}_{1} \; q^{(1)}_{\mu\nu}+p^{}_{2} \; q^{(2)}_{\mu\nu} \;.
\end{equation}
The number $m^{(\rho)}$ of elementary events, generated in measurement $\rho$, is Poisson distributed,
\begin{eqnarray}\label{eq:Poisson}
{\cal P}(m^{(\rho)}\CS \lambda_\rho) = \frac{(\lambda_\rho)^{m^{(\rho)}}}{m^{(\rho)}!}\;e^{-\lambda_\rho}\;.
\end{eqnarray}
In the $\beta$ measurement $\lambda_\beta = \lambda_1 + \lambda_2$ and the $m^{(\beta)}$ events are binomially distributed between channels 1 and 2. The probability 
to have $m^{(j)}$ events of channel $j$, out of a total of $m^{(\beta)}$ events, is therefore
\begin{equation}
{\cal B}(m^{(j)}|m^{(\beta)},p_{j}) = {m^{(\beta)}\choose m^{(j)}} (p_{j})^{m^{(j)}} (1-p_{j})^{m^{(\beta)}-m^{(j)}}\;.
\end{equation}
Then the probability to find $m^{(j)}$ events of channel $j$ in one measurement is
\begin{widetext}
\begin{eqnarray}
 P(m^{(j)}\CS \lambda_\beta,p_j,\BK) &= \sum_{m^{(\beta)}=0}^{\infty}
P(m^{(j)}\CS m^{(\beta)},p_j,\xxx{\lambda_\beta},\BK)P(m^{(\beta)}\CS \xxx{p_j},\lambda_\beta,\BK) ={\cal P}(m^{(j)}\CS \lambda_\beta p_{j} )\;.
\end{eqnarray}
\end{widetext}
Irrelevant parameters behind the conditional bar have been crossed out.
The first factor is binomial and the second factor is Poisson distributed.
The result is, obviously, also a Poisson distribution with mean $\lambda_{j}=\lambda_\beta p_{j}$. We therefore have equivalently, since $\lambda_{\beta}=\lambda_{1}+\lambda_{2}$,
\begin{equation}\label{eq:lambda}
  p_{j} = \frac{\lambda_{j}}{\lambda_{1}+\lambda_{2}}\;.
\end{equation}
Let  $N\Ro_{\mu\nu}\RR$ be the number of events  $(\mu\nu)$ in measurement $\rho$ and   $\{N\Ro_{\mu\nu}\RR\}$ the set of counts for all pairs $(\mu,\nu)$.
For better readability, we will denote a set of the form $\{s^{(x)}_{\mu\nu}\}$ simply by $s^{(x)}$; e.g., $N\Ro = \{N\Ro_{\mu\nu}\RR\}$. 
Then the joint probability for these counts is
\begin{widetext}
\begin{equation}
 P(N\Ro\CS q\Ro,\lambda_\rho,\BK) =\sum_{m^{(\rho)}=0}^{\infty}
P(N\Ro\CS m^{(\rho)},q\Ro,\xxx{\lambda_\rho},\BK)
P(m^{(\rho)}\CS\xxx{q\Ro},\lambda_\rho,\BK)\;.
\end{equation}
\end{widetext}
The first factor is a multinomial and the second a Poisson distribution. \new{In total we obtain 
\begin{eqnarray}\label{eq:joint:Poisson}
P(N\Ro\CS q\Ro,\lambda_\rho,\BK) &= \prod_{\mu\nu} {\cal P}(N^{(\rho)}_{\mu\nu}\CS \lambda_\rho q\Ro_{\mu\nu})\;,
\end{eqnarray}
which states that each count $N^{(\rho)}_{\mu\nu}$ is independently Poisson distributed with its individual mean $\lambda_\rho q^{(\rho)}_{\mu\nu}$.} For the two measurements we have to use the
corresponding probabilities given in \eq{eq:q:alpha} and
\eq{eq:q:beta}.
The mean values of the Poisson distributions for the two measurements are therefore
\begin{align}
\langle N^{(\alpha)}_{\mu\nu}  \rangle =& \lambda_1 q^{(1)}_{\mu\nu} \notag \\
\langle N^{(\beta)}_{\mu\nu}  \rangle =& 
\lambda_{\beta} q_{\mu\nu}^{(\beta)} \notag \\
 =&(\lambda_1+\lambda_2) 
\big(p_{1} q^{(1)}_{\mu\nu}+p_{2} q^{(2)}_{\mu\nu}\big)\notag \\
 =& \lambda_{1} q^{(1)}_{\mu\nu} + \lambda_{2} q^{(2)}_{\mu\nu}\;.
\end{align}
Hence, one is prompted to simply  subtract the counts of coincidence measurements 
to get rid of the background signal $(\lambda_{1}q_{\mu\nu}^{(1)})$ in measurement $\beta$.
The  difference $\Delta N_{\mu\nu} :=  N^{(\beta)}_{\mu\nu}-N^{(\alpha)}_{\mu\nu}$ of two Poisson numbers 
obeys the \new{Skellam distribution \cite{skellam_1946}}, with mean and variance resulting in
\begin{align}
\langle \Delta N_{\mu\nu} \rangle&= 
\langle N^{(\beta)}_{\mu\nu} \rangle - \langle N^{(\alpha)}_{\mu\nu}  \rangle=
\lambda_{2} q^{(2)}_{\mu\nu} \;, \notag \\
\langle \big(\Delta N_{\mu\nu}\big)^{2} \rangle&= 
\langle N^{(\beta)}_{\mu\nu} \rangle + \langle N^{(\alpha)}_{\mu\nu}  \rangle     \;.
\end{align} 
Obviously, the difference of the counts is an unbiased estimator
of the sought-for quantity $q^{(2)}_{\mu\nu}$. However, in regions
of the spectrum, where the background  dominates, i.e., $\lambda_1 q^{(1)}_{\mu\nu}\gg \lambda_2 q^{(2)}_{\mu\nu} $, the variance will in general be much greater than the difference of the counts, which will usually lead to nonphysical negative results and large uncertainties. 

So far, we have exploited all events, detected during the individual measurements.
Now we turn to the single coincidence evaluation, where only those measurements are 
recorded, in which exactly one electron-ion pair $(\mu,\nu)$ has been detected.
As we will see below, using only the single coincidence events out of the measured data influences the statistics. There is a further problem, that needs to be addressed: The detectors are not perfect, which results in false coincidences.
To begin with, we will ignore this problem, and consider the single coincidences detection method  for perfect detectors. The probability that exactly one event, with indices $(\mu\nu)$ say, happens can be computed from \eq{eq:joint:Poisson}, by setting all $N^{(\rho)}_{\mu'\nu'}=0$, except for $N^{(\rho)}_{\mu\nu}=1$.
This leads to
\begin{align}\label{eq:}
 P^{(\rho)}_{\mu\nu} &:=\bigg( \lambda_\rho q_{\mu\nu}^{(\rho)} e^{-\lambda_\rho q_{\mu\nu}^{(\rho)}}\bigg) \;
\prod_{\mu'\nu'}^{\ne \mu\nu} e^{-\lambda_\rho q_{\mu'\nu'}^{(\rho)}} \notag \\
&= \lambda_\rho q_{\mu\nu}^{(\rho)} \;
 e^{-\lambda_\rho \sum_{\mu'\nu'} q_{\mu'\nu'}^{(\rho)}} \notag \\
&=\lambda_\rho q_{\mu\nu}^{(\rho)} e^{-\lambda_\rho} \;.
\end{align}
In this type of coincidence measurement, there are two possible outcomes:
A coincidence is detected with probability $P^{(\rho)}_{\mu\nu}$ or not detected with the complementary probability $1-P^{(\rho)}_{\mu\nu}$. The latter case covers the cases that there was no coincidence or more than one.
It is therefore a Bernoulli type of experiment and the probability that the number of single coincidences is $n^{(\rho)}_{\mu\nu}$ 
given a total of ${\cal N}_{p}$ measurements is binomial,
\begin{equation}\label{eq:}
P(n^{(\rho)}_{\mu\nu}\CS N_{p}, P^{(\rho)}_{\mu\nu},\BK) =
{\cal B}(n^{(\rho)}_{\mu\nu}\CS {\cal N}^{(\rho)}_{p}, P^{(\rho)}_{\mu\nu})\;.
\end{equation}
The expectation value of the weighted difference between the counts, measured in experiment $\beta$ and experiment $\alpha$, with equal numbers of measurements in both experiments, is given by
\begin{align}
 \left< n^{(\beta)}_{\mu\nu} - \chi n^{(\alpha)}_{\mu\nu} \right> = 
 {\cal N}_{p} \bigg( P^{(\beta)}_{\mu\nu}- \chi P^{(\alpha)}_{\mu\nu} \bigg) \notag \\
  =  {\cal N}_{p} e^{-\lambda_1} \left( \lambda_1 q^{(1)}_{\mu\nu} \left( e^{-\lambda_2 } - \chi \right) + \lambda_2 q^{(2)}_{\mu\nu} e^{-\lambda_2 }\right) \;.
\end{align}
The simple subtraction, namely setting $\chi = 1$, is not an unbiased estimator for $q^{(2)}_{\mu\nu}$ anymore and would lead to erroneous results.
By choosing the weight $\chi = e^{-\lambda_2}$, this can be overcome.
But, as pointed out before, this weighted subtraction does not take into account 
false coincidences due to imperfect detectors and can lead to nonphysical negative results and large uncertainties.

\new{Summarized, we have the following data analysis problem:
\begin{enumerate}
    \item In the measurement, ions and electrons are created in pairs in the two independent channels $1$ and $2$. The number of electron-ion pairs in each channel follows a Poissonian distribution with the parameters $\lambda_1$ and $\lambda_2$.
    \item Furthermore, each electron-ion pair is assigned with an electron energy $\nu$ and an ion mass $\mu$ according to a multinomial distribution containing the spectra $q^{(1)}_{\mu\nu}$ and $q^{(2)}_{\mu\nu}$ we want to determine.
    \item Experimentally we can measure channel $1$ in the pump-only measurement or the combination of channels $1$ and $2$ in the pump-probe measurement.
    \item To reconstruct the spectra we need the connection between ions and electrons and therefore use only single coincidence measurements, where exactly one electron-ion pair is detected.
    \item Due to imperfect detectors with a detection probability less than unity false coincidences arise in the coincidence method.
\end{enumerate}}
A powerful and, as a matter of fact, the only consistent approach to take all experimental features,  uncertainties, and additional prior knowledge into account is provided by Bayesian probability theory \cite{jaynes_probability_2003,linden_bayesian_2014}.

\section{Bayesian data analysis}
\label{sec:bayes}

\new{We will now use Bayesian probability theory to calculate the {PDF} for $q^{(2)}=\{q^{(2)}_{\mu\nu}\}$, the spectrum of channel 2, given the measured dataset $D_{1}$, which contains the count rates $n^{(\alpha)}=\{n^{(\alpha)}_{\mu\nu}\}$ and $n^{(\beta)}=\{n^{(\beta)}_{\mu\nu}\}$.}
As set out  in the previous section, $n^{(\rho)}_{\mu\nu}$ counts how often the pair  $(E_{\nu},M_{\mu})$ was detected as single coincidence event
during the experiment $\rho$. But in contrast  to the previous section,
there may be false coincidences involved.
In this paper we only use single coincidence events for estimating $q^{(2)}$, which is justified by the fact, that especially these events include relevant information about the spectrum. The case of detecting more than one electron-ion pair does not allow us to link an electron to the ion it originates from and the Bayesian approach would  be  different.
In addition to $D_{1}$, we also use a second dataset $D_{2}$ containing
$N^{(\alpha)}_{N_{e},N_{i}}$ and $N^{(\beta)}_{N_{e},N_{i}}$, which counts how many measurements lead to the detection of $N_e$ electrons and $N_i$ ions during the experiments $\alpha$ and $\beta$, respectively. 
In this case, it is expedient to use all detected events, not just
single coincidences.
More details will be given below. This dataset will be used to determine the unknown parameters $\pi:=\{\lambda_1, \lambda_2, \xi_i, \xi_e\}$. 
$\lambda_1$ and $\lambda_2$ were already introduced in the previous section, and $\xi_i$ and $\xi_e$ are the detection probabilities for ions and electrons, respectively.
In a first step we introduce the unknown parameters $\pi$ using the marginalization rule of Bayesian probability theory,
\begin{align}
p(q^{(2)}\CS D_{1},D_{2}, \BK) =
 \int d\pi & \;p(q^{(2)}\CS D_{1},D_{2}, \oBK) \notag \\ & \;\times  p(\pi\CS D_{1},D_{2}, \BK)
 \label{eq:pr_q2:a}\;.
\end{align}
\new{The integration over $\pi$ means integrating out each parameter included in $\pi$. The domain of each integration parameter, and therefore the integration region, should be clear from the context.
We keep this abbreviated notation during the whole derivation.}
The dataset $D_{2}$ contains no detailed information concerning electron energy and ion mass and can therefore be omitted in the first factor. Similarly, in the second factor, the dataset $D_{1}$ carries negligible information about the parameters $\lambda_{j}$ and $\xi_{e}$, $\xi_{i}$, and will be suppressed as well. Then, therefore, we have
\begin{align}
p(q^{(2)}\CS D_{1},D_{2}, \BK) =
 \int d\pi & \; p(q^{(2)}\CS D_{1}, \oBK) \notag \\ & \; \times p(\pi\CS D_{2}, \BK)
 \label{eq:pr_q2}\;.
\end{align}
There appear two new probability distributions, one for $q^{(2)}$ given the measured coincidences $D_{1}$ and the parameters \new{$\pi$} and one for the parameters $\pi$ given the measurements $D_{2}$.

\subsection{Reconstructing the spectrum $q^{(2)}$\label{sec:spectra}}

The first \new{PDF} in \eq{eq:pr_q2} can be calculated using again the marginalization rule to introduce $q^{(\alpha)}$ and $q^{(\beta)}$,
\begin{widetext}
\begin{align} \label{eq:q2_nanb}
 p(q^{(2)}\CS n^{(\alpha)}, n^{(\beta)},\oBK) = \int dq^{(\alpha)} dq^{(\beta)}\;
 p(q^{(2)}\CS q^{(\alpha)},q^{(\beta)},\oBK) \;p(q^{(\alpha)}\CS n^{(\alpha)},\oBK) p(q^{(\beta)}\CS n^{(\beta)},\oBK)\;,
\end{align}
\end{widetext}
with $dq^{(\rho)} = \prod_{\mu\nu} dq^{(\rho)}_{\mu\nu}$. 
In the last two factors, which represent $p(q^{(\alpha)},q^{(\beta)}\CS n^{(\alpha)},n^{(\beta)},\oBK)$, 
we have exploited the fact that the two experiments $\alpha$ and $\beta$ are not correlated.
Knowing the spectra $q^{(\alpha)}$ and $q^{(\beta)}$, the spectrum of channel $2$ is uniquely determined  due to \eq{eq:q:beta} and \eq{eq:lambda}, resulting in
\new{\begin{widetext}
\begin{align}
p(q^{(2)}\CS q^{(\alpha)},q^{(\beta)},\oBK) =
\delta\left( q^{(2)}  -  \bigg[\frac{\lambda_1 + \lambda_2} {\lambda_2}\;q^{(\beta)}  -  \frac{\lambda_1}{\lambda_2} q^{{(\alpha)}}\bigg] \right)
=\bigg(\frac{\lambda_{2}}{\lambda_{1}+\lambda_{2}}\bigg)^{\cal N}\;\delta\left( q^{(\beta)}  -  
p_1 q^{(\alpha)} - p_2 q^{(2)} \right) \;,
\end{align}
\end{widetext}}
with ${\cal N}={\cal N}_{\mu} {\cal N}_{\nu}$, {where ${\cal N}_\mu$ and ${\cal N}_\nu$ are the total numbers of bins of $\{ M_\mu \}$ and $\{ N_\nu \} $, respectively.}
Hence, we can readily integrate out  $q^{(\beta)}$.
For a better readability, the superscript $\rho$ will be omitted in the following considerations. 
It can easily be included again at the end.
At this point it is crucial to  recall that $q_{\mu\nu}\RR$ is the probability that an electron with energy $E_{\nu}$ and an ion of mass $M_{\mu}$ are created in an elementary event (ionization of one molecule). The link to the experimental observations, however, is $\tilde q^{(\rho)}_{\mu\nu}$, the probability that -- given a single coincidence is detected -- the measured electron has energy $E_{\nu}$ and the detected ion has mass $M_{\mu}$. 
\new{There is a simple relation between the two probabilities $q \RR_{\mu\nu}$ and $\tilde q \RR_{\mu\nu}$}, which is derived in Appendix  
\ref{app:a}
\begin{equation}\label{eq:q:tilde:q}
  \tilde q \RR_{\mu\nu} =\frac{ q \RR_{\mu\nu} 
+ \kappa \RR\; q_{\cdot\nu} q_{\mu\cdot} }{1 +  \kappa \RR}\;,
\end{equation}
with $\kappa \RR=\lambda \RR \bar\xi_{e} \bar\xi_{i}$, where $\bar \xi_{e} = (1-\xi_{e})$ and $\bar \xi_{i} = (1 - \xi_{i})$\new{, and the marginal probabilities $q_{\mu\cdot}=\sum_{\nu'}q_{\mu\nu'}$ and $q_{\cdot\nu}=\sum_{\mu'}q_{\mu'\nu}$}.
We abbreviate this bijection by $\tilde q \RR = \tilde Q \RR(q \RR)$.
The interpretation is quite intuitive:
The false coincidencies are represented by the term $\kappa \RR\; q_{\cdot\nu} q_{\mu\cdot}$. False coincidences 
require that electron and ion detection fails, which explains the factor  $\bar \xi_{i} \bar \xi_{e}$.
If the detectors are perfect, $\kappa$ becomes zero and there are no false coincidencies and $\tilde q_{\mu\nu} = q_{\mu\nu}$ holds.
The additional factor $\lambda$ is due to the fact that at least a second elementary event is needed to observe false coincidences. 
In the case of a false coincidence event corresponding to $(\mu\nu)$, an electron with energy $E_{\nu}$ is required, for which the  probability  is given by $q_{\cdot\nu}$. This marginal probability corresponds to elementary events with electron energy  $E_{\nu}$ and any mass $M_{\mu'}$. Similarly the probability for detecting a mass $M_{\mu}$, irrespective of the electron energy associated with  the elementary event,
is given by $q_{\mu\cdot}$. This explains the factor $q_{\cdot\nu}q_{\mu\cdot}$.
The denominator $1+\kappa$ is required for the normalization of $\tilde q_{\mu\nu}$.
Summation over $\mu$ or $\nu$ reveals that the marginal probabilities are identical, i.e., $\tilde q_{\mu\cdot} = q_{\mu\cdot}$ and
 $\tilde q_{\cdot\nu} = q_{\cdot\nu}$. This is very reasonable,
 as the probability distribution of the measured  electron 
 energies  $E_{\nu}$ is the same as the electron 
 energy distribution in the elementary events, because
it does not depend on the correct or false assignment of corresponding masses.

 The \new{PDF}  $p(\tilde q \RR \CS n \RR ,\oBK)$ can easily be calculated using Bayes' theorem,
\begin{equation}
 p(\tilde q \RR \CS n \RR ,\oBK) = \frac{1}{Z}\;P( n \RR \CS  \tilde q \RR,\oBK) \;p(  \tilde q \RR\CS\xxx{\pi},\BK) \;.
\end{equation}
According to the Appendix \ref{app:c} the likelihood function $P( n \RR \CS  \tilde q \RR,\oBK) $ is multinomial.
We  use a \new{Dirichlet prior \cite{linden_bayesian_2014} }for $\tilde q \RR$
\begin{equation}\label{eq:dirichlet}
p(\tilde q \RR \CS \BK) = \frac{1}{B(\{c_{\mu\nu}\})} 
\prod_{\mu\nu}  \tilde q \RR_{\mu\nu}^{c \RR_{\mu\nu}-1}\;\delta(\tilde S \RR -1)\;,
\end{equation}
with 
$\tilde S \RR=\sum_{\mu\nu} \tilde q \RR_{\mu\nu}$, and  
the normalization $B(\{c_{\mu\nu}\})$ being the multivariate beta function.
We can always choose the prior to be uninformative (flat), by setting all $c \RR_{\mu\nu} = 1$. The posterior is a Dirichlet \new{PDF} as well. 
\begin{equation}\label{eq:posterior:tilde:q}
 p(\tilde q \RR \CS n \RR ,\oBK) =
 \frac{1}{B\left( \left\{ n \RR_{\mu\nu}+c \RR_{\mu\nu}\right\} \right)}
 \prod_{\mu\nu} \tilde q_{\mu\nu}^{n_{\mu\nu}+c_{\mu\nu}-1}\;\delta(\tilde S \RR-1)\;.
\end{equation}
Based on  the usual transformation rule for \new{PDF}s, 
\begin{equation}\label{eq:q:n}
p(q \RR \CS n \RR ,\oBK) =  
p(\tilde q \RR \CS n \RR ,\oBK) \bigg|\frac{d \tilde Q(q)}{d q}\bigg| \;,
\end{equation}
we show in Appendix \ref{app:b}  that \eq{eq:q:n} eventually becomes
\begin{widetext}
\begin{equation}
 p( q^{(\rho)}\CS n^{(\rho)},\oBK) 
 =  \frac{\big( 1+\kappa_{\rho} \big)^{-({\cal N}_{\mu}-1)( {\cal N}_{\nu}-1)}
}{B\left( \left\{ n^{(\rho)}_{\mu\nu}+c^{(\rho)}_{\mu\nu}\right\} \right)} \prod_{\mu\nu} 
  \big(\tilde Q^{}_{\mu\nu}(q^{(\rho)})\big)^{n^{(\rho)}_{\mu\nu}+c^{(\rho)}_{\mu\nu}-1} \delta\left(\sum_{\mu\nu}q_{\mu\nu}^{(\rho)}-1\right)\;.
\end{equation}
\end{widetext}
Here we have reintroduced the superscript $\rho$ and defined $\kappa_{\rho}=\lambda_{\rho} \oxi_{e}\oxi_{i}$.
Finally, the sought probability distribution in \eq{eq:q2_nanb} becomes
\new{\begin{widetext}
\begin{eqnarray}
 p(q^{(2)}\CS n^{(\alpha)}, n^{(\beta)},\oBK) = \bigg(\frac{\lambda_{2}}{\lambda_{1}+\lambda_{2}}\bigg) ^{\cal N} \int 
  dq^{(\alpha)}dq^{(\beta)}  \;p(q^{(\alpha)}\CS n^{(\alpha)},\oBK) \;p(q^{(\beta)}\CS n^{(\beta)},\oBK)
\delta\left( q^{(\beta)}  -  
p_1 q^{(\alpha)} - p_2 q^{(2)} \right) \;.
\end{eqnarray}
\end{widetext}}
The integral is a convolution, which describes the subtraction of the two spectra $q^{(\alpha)}$ and $q^{(\beta)}$.

\subsection{The \new{PDF} for  the parameters $\lambda_1$, $\lambda_2$, $\xi_i$ and $\xi_e$\label{sec:parameter}}

We still need to determine the \new{PDF}  $p(\pi \CS D_{2}, \BK)$ in \eq{eq:pr_q2}.
For estimating the parameters $\pi=\{\lambda_1,\lambda_2, \xi_i, \xi_e\}$ we will use the second dataset $D_{2}$. The superscript $\rho$ will be omitted in the following considerations.
The dataset $D_{2}$ contains $\{ N_{N_e,N_i} \} $, the total numbers of measurements, in which $N_e$ electrons and $N_i$ ions were detected, irrespective of electron energy or ion mass.
For instance, $N_{1,1}$ is the total number of single coincidences measured in the experiment, which consists of ${\cal N}_{p}$ measurements.  
The sum over all $\{ N_{N_e,N_i} \}$ is the number of measurements ${\cal N}_{p}$,
\begin{equation}
 {\cal N}_{p} = \sum_{N_e,N_i} N_{N_e,N_i} \;.
\end{equation}
For better readability we introduce a more compact notation. We enumerate the possible count-pairs
$(N_{e},N_{i})$ by an integer $l$, where $l=0$ stands for the pair $(0,0)$,  $l=1$ for $(0,1)$, $l=2$ for $(1,0)$, and so on. Then $\{N_{N_{e},N_{i}}\}\to \{N_{l}\}$. Moreover, we measure $\{N_{l}\}$
in the pump-only ($\alpha$) and the pump-probe ($\beta$) experiment, denoted by 
$N^{(\rho)}=\{N^{(\rho)}_{l}\}$.
Now we can proceed with Bayes' theorem,
{
\begin{align}
 p(\pi \CS N^{(\alpha)}, N^{(\beta)}, \BK) =&
\frac{1}{Z}\;P(N^{(\alpha)}, N^{(\beta)} \CS \pi,\BK)\;
p(\pi \CS  \BK)\label{eq:pr:parameters} \notag \\
=&\frac{1}{Z}\;P(N^{(\alpha)} \CS \lambda_1, {\lambda_2}, \xi_i, \xi_e, \BK)  \notag \\
 & \; \times P(N^{(\beta)} \CS \lambda_1, \lambda_2, \xi_i, \xi_e, \BK)\;  \notag \\
 & \times p(\lambda_1, \lambda_2, \xi_i, \xi_e \CS  \BK)\;.
\end{align}}
In the last step we have exploited the fact that the two experiments are uncorrelated.
For real application we can use uniform priors for $\xi_e$ and $\xi_i$ and  \new{Jeffreys' prior \cite{Jeffreys_Prior_1946}} for $\lambda_{j}$, i.e.,
$
p(\lambda_1, \lambda_2, \xi_i, \xi_e \CS  \BK) \propto \frac{1}{\lambda_{1}\lambda_{2}}
$, resulting in 
\begin{align}\label{eq:parameters}
 p(\pi\CS N^{(\alpha)}, N^{(\beta)}, \BK)
=&\frac{1}{Z'}\;P(N^{(\alpha)} \CS \lambda_{\alpha}, \xi_i, \xi_e, \BK)  \notag \\
\;& \times P(N^{(\beta)} \CS \lambda_{\beta}, \xi_i, \xi_e, \BK)\;
\frac{1}{\lambda_{1}\lambda_{2}}\;,
\end{align}
with $\lambda_{\alpha}=\lambda_{1}$, and $\lambda_{\beta}=\lambda_1+ \lambda_2$.
Next, we consider the likelihood $P(N^{(\rho)} \CS \lambda_\rho, \xi_i, \xi_e, \BK)$ term. 
In Appendix \ref{app:Na} we have computed the probability $P_{l} = P_{N_{e}N_{i}}$ that in one measurement the count-pair is $(N_{e}, N_{i})$.  This probability actually depends on $\rho$ via $\lambda_{\rho}$.
Clearly, the probability  $P(N^{(\rho)} \CS \lambda_\rho, \xi_i, \xi_e, \BK)$ is multinomial, 
\begin{equation}\label{eq:Na:lambda}
 p( N^{(\rho)} \CS \lambda_{\rho},\xi_e,\xi_i, \BK) = \frac{{\cal N}^{(\rho)}_{p}!}{\prod_{l=0}^{l^{*}} N^{(\rho)}_{l}!} \prod_{l=0}^{l^{*}} \big(P_l^{(\rho)}\big)^{N_l\Ro}\;.
\end{equation}
In principle, $l^{*}=\infty$, but, as argued before, it is expedient to adjust the experiment such that $\lambda$  is $O(1)$. Then   $P^{(\rho)}_{l}$ rapidly decreases with $l$ (see Appendix \ref{app:Na}) and it suffices to restrict to $l< l^{*}$, with a moderate value for $l^{*}$.
All other events, belonging to $l\ge l^{*}$ are combined in one auxiliary event $l^{*}$, with 
\begin{equation}
N\Ro_{l^{*}}={\cal N}^{(\rho)}_{p}-\sum_{l=0}^{l^{*}-1}N\Ro_{l}\;,\qquad
P\Ro_{l^{*}}=1-\sum_{l=0}^{l^{*}-1} P\Ro_{l}\;.
\end{equation}

\subsection{Evaluating the probability density for $q^{(2)}$}

We now have determined the probability distribution $p(q^{(2)}\CS D_1, D_2, \BK)$.
The result can be summarized as following. 
\new{
\begin{widetext}
\begin{center}
\begin{mdframed}
\parbox{1.0\linewidth}{\vspace{0.3cm}\centerline{\new{Summary of the definitions and the derived probabilities.}}\rule{\linewidth}{0.4pt}}
{
\begin{center}
\begin{minipage}{16cm}\new{
Reconstruction of the spectra of channels $1$ and $2$ out of a pump-only ($\alpha$) experiment producing only the spectrum of channel $1$ and of a pump-probe ($\beta$) experiment producing a mixture of the spectra of channels $1$ and $2$. The important variables are:}
\end{minipage}
\end{center}
\begin{center}
\begin{minipage}{14cm}\new{
\begin{itemize}
\item[$q^{(x)}$ \dots]  Spectra of the channels, respectively, experiments, $x\in\{1,2,\alpha,\beta\}$. $q^{(x)}=\{q^{(x)}_{\mu\nu}\}$, where $\mu$ depicts the measured ion masses and $\nu$ the electron energies.
${\cal N}_\mu$ and ${\cal N}_\nu$ are the numbers of elements in $\{ \mu \}$ and $\{ \nu \} $, respectively, and ${\cal N}={\cal N}_{\mu} {\cal N}_{\nu}$.
\item[$\pi$ \dots]  Summarizes the parameter $\lambda_1$, $\lambda_2$, $\xi_i$, and $\xi_e$, where $\lambda_1$ and $\lambda_2$ are parameters of the Poisson distributions determining the measured count rates and $\xi_i$ and $\xi_e$ are the detection probabilities of ions and electrons, respectively.
\item[$D_1$ \dots]  Dataset containing the count rates 
$n^{(\rho)}=\{n^{(\rho)}_{\mu\nu}\}$, $\rho\in\{\alpha,\beta\}$, of the coincidence events.
\item[$D_2$ \dots]  Dataset containing the total numbers of measurements $N^{(\rho)}=\{N^{(\rho)}_{N_{e},N_{i}}\}$ in which $N_e$ electrons and $N_i$ ions were detected.
\end{itemize}}
\end{minipage}
\end{center}
\begin{center}
\begin{minipage}{16cm}\new{
The probability for the spectra $q^{(2)}$ is:}
\end{minipage}
\end{center}\new{
\begin{align}
 p(q^{(2)}\CS D_1, D_2, \BK) &= \int d\pi \; p(q^{(2)}\CS D_1, \pi, \BK) p(\pi \CS D_2, \BK) \;. \notag
 \end{align}}
 \begin{center}
\begin{minipage}{16cm}\new{
For the first probability distribution we derived in Sec. \ref{sec:spectra}:}
\end{minipage}
\end{center}\new{
 \begin{align}
 p(q^{(2)}\CS D_1, \pi, \BK) &= \bigg(\frac{\lambda_{2}}{\lambda_{1}+\lambda_{2}}\bigg) ^{\cal N} \int 
  dq^{(\alpha)}dq^{(\beta)} \;p(q^{(\alpha)}\CS n^{(\alpha)},\oBK) \;p(q^{(\beta)}\CS n^{(\beta)},\oBK) \delta\left( q^{(\beta)}  -  
p_1 q^{(\alpha)} - p_2 q^{(2)} \right) \notag\\
   &\text{with~~} p( q^{(\rho)}\CS n^{(\rho)},\oBK) =  \frac{\big( 1+\kappa_{\rho} \big)^{-({\cal N}_{\mu}-1)( {\cal N}_{\nu}-1)}
}{B\left( \left\{ n^{(\rho)}_{\mu\nu}+c^{(\rho)}_{\mu\nu}\right\} \right)}  \prod_{\mu\nu} 
  \big(\tilde Q^{}_{\mu\nu}(q^{(\rho)})\big)^{n^{(\rho)}_{\mu\nu}+c^{(\rho)}_{\mu\nu}-1} \delta\left(\sum_{\mu\nu}q_{\mu\nu}^{(\rho)}-1\right) \;, \notag\\
  & p_i = \frac{\lambda_i}{\lambda_1+\lambda_2} \; , \;
   \tilde Q \RR_{\mu\nu}(q^{(\rho)}) =\frac{ q^{(\rho)} \RR_{\mu\nu} 
+ \kappa_\rho \RR\; q^{(\rho)}_{\cdot\nu} q^{(\rho)}_{\mu\cdot} }{1 +  \kappa_\rho \RR} \text{~~and~~} \kappa_\rho = \lambda_\rho(1-\xi_{e}) (1-\xi_{i}) \;. \notag
\end{align}}
 \begin{center}
\begin{minipage}{16cm}\new{
For the second probability distribution we derived in Sec. \ref{sec:parameter},}
\end{minipage}
\end{center}\new{
 \begin{align}
 p(\pi \CS D_2, \BK) &\propto \;P(N^{(\alpha)} \CS \lambda_{\alpha}, \xi_i, \xi_e, \BK)
\;P(N^{(\beta)} \CS \lambda_{\beta}, \xi_i, \xi_e, \BK)\;
\frac{1}{\lambda_{1}\lambda_{2}}\; \notag\\
 &\text{with~~} p( N^{(\rho)} \CS \lambda_{\rho},\xi_e,\xi_i, \BK) = \frac{{\cal N}^{(\rho)}_{p}!}{\prod_{l=0}^{l^{*}} N^{(\rho)}_{l}!} \prod_{l=0}^{l^{*}} \big(P_l^{(\rho)}\big)^{N_l\Ro} \;, \; \lambda_\alpha = \lambda_1 \text{~~and~~} \lambda_\beta = \lambda_1+\lambda_2 \;. \notag \\
 &\text{For the explanation of ${\cal N}^{(\rho)}_{p}$, $N_l\Ro$ and $P_l^{(\rho)}$, see Sec. \ref{sec:parameter}.} \notag
\end{align}}
}
\end{mdframed}
\end{center}
\end{widetext}
}

A suitable technique for sampling from a probability distribution is Markov Chain Monte Carlo (MCMC), which is based on constructing a Markov chain that has the desired distribution as its equilibrium distribution. \new{The technique is standard in Bayesian probability theory \cite{linden_bayesian_2014} (and references therein)}.
\new{In particular, we are interested in the mean and the variance of $q^{(2)}$.
But we could as well determine the expectation value of an arbitrary function of $\mathcal{O}(q^{(2)})$ by evaluating the integral}
\begin{equation}
\left< \mathcal{O}(q^{(2)}) \right> = \int  dq^{(2)}\mathcal{O}(q^{(2)})p(q^{(2)}\CS D_1, D_2, \BK) \;.
\end{equation}
\new{Even more general expectation values} depending  on the parameters $\theta = \{q^{(\alpha)}, q^{(2)}, \lambda_1, \lambda_2, \xi_i, \xi_e\} $ can be calculated using
\begin{equation}
\left< \mathcal{O}(\theta) \right> = \int d\theta \mathcal{O}(\theta) f(\theta)\;,
\label{eq:observable}
\end{equation}
with
\new{\begin{widetext}
\begin{equation}
 f(\theta) = \bigg(\frac{\lambda_{2}}{\lambda_{1}+\lambda_{2}}\bigg) ^{\cal N} \int 
  dq^{(\beta)}  p(q^{(\alpha)}\CS n^{(\alpha)},\oBK) \;p(q^{(\beta)}\CS n^{(\beta)},\oBK) \delta\left( q^{(\beta)}  -  
p_1 q^{(\alpha)} - p_2 q^{(2)} \right) p(\pi\CS D_2, \BK) \;.
\end{equation}
\end{widetext}}
We have used the Metropolis Hastings algorithm to generate the Markov chain $\{ \theta^k \} $.
We start with a parameter set $\theta^{k=1}$ and every new parameter set $k+1$ can be proposed by varying parameters in the old parameter set $k$. The  new parameter set $k+1$ is accepted with the probability
\begin{equation}
 P_{\text{acc}} = \min\left\{ 1  ,\frac{f(\theta^{k+1})}{f(\theta^{k})} \right\} \;.
\end{equation}
\new{It occurs that the first $10$-$20$~\% of a Markov chain have to be discarded to ensure that the rest of the Markov chain is independent of the initial state $\theta^{k=1}$, and therefore the Markov chain is thermalized to the desired distribution.}
For calculating confidence intervals for expectation values, such as that in equation \eq{eq:observable}, the states in the Markov chain have to be uncorrelated, which can be ensured by taking only every $N_{\text{run}}$th state of the Markov chain.  $N_{\text{run}}$ can be controlled by evaluating the autocorrelation function or using techniques like binning and jackknife.
Finally, the observable can be estimated by
\begin{equation} 
 \mathcal{O} := \left< \mathcal{O}(\theta) \right> \approx \frac{1}{N_{\text{Markov}}} \sum_k \mathcal{O}(\theta^{k}) \;.
\end{equation}
\new{The confidence intervals are estimated from
\begin{equation}
\Delta \mathcal{O}:= \frac{\sigma_{\mathcal{O}}}{\sqrt{N_{\text{Markov}}}}\;,
\end{equation}
for which it is crucial that the $N_{\text{Markov}}$ elements of the Markov chain are uncorrelated. The variance 
\begin{equation}
\sigma_{\mathcal{O}}^{2}= \left<\mathcal{O}(\theta)^{2} \right> - \left<\mathcal{O}(\theta) \right>^{2} 
\end{equation} 
can in turn be estimated from the Markov chain.} Alternatively, the uncertainty 
$\Delta \mathcal{O}$ can be determined from independent MCMC runs.

\section{Mock data analysis}
\label{sec:mock}

\begin{figure*}
    \centering
    \includegraphics[width=1.9\columnwidth,angle=0]{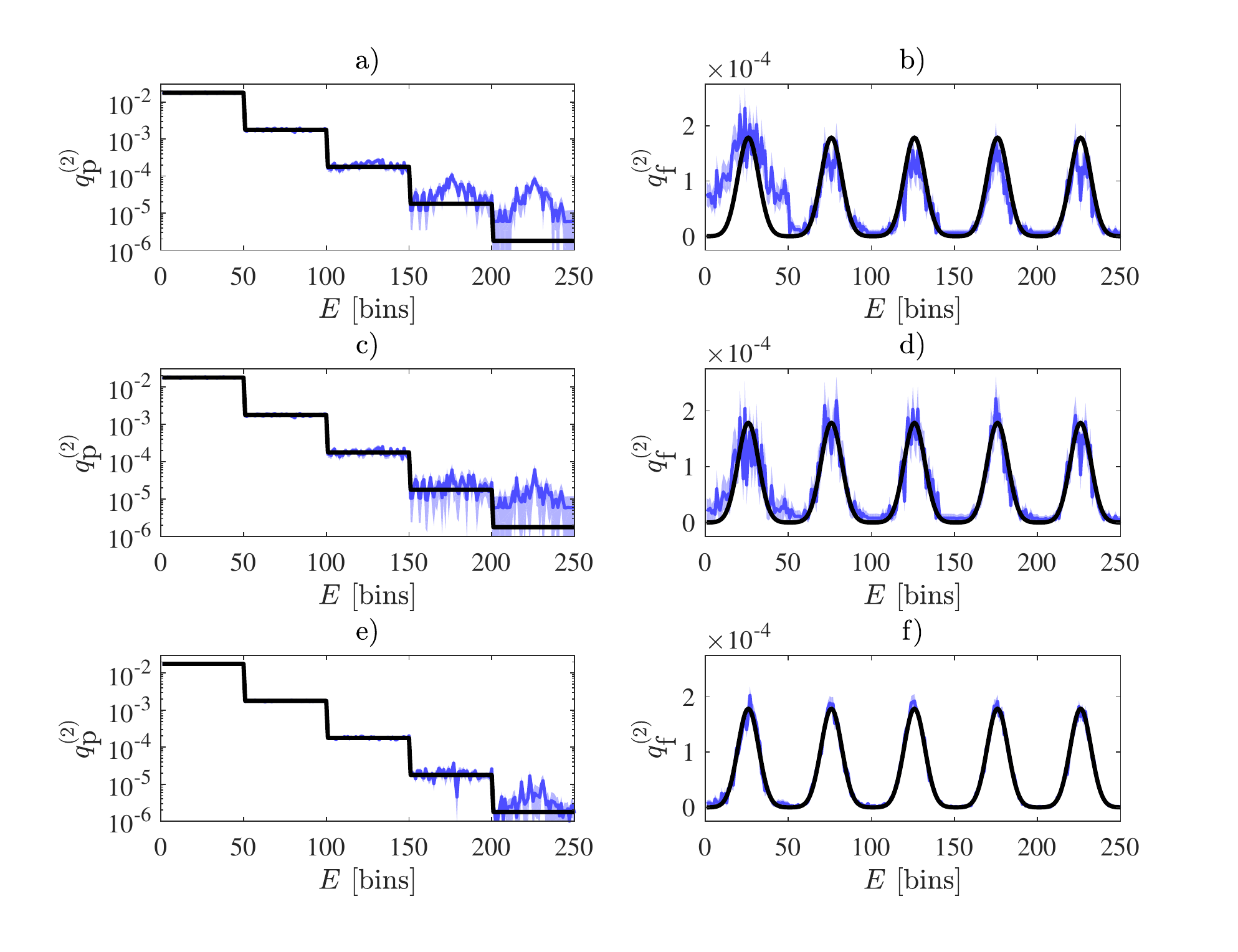}
    \caption{Test calculation with mock data. The black lines are the spectra used to generate the data (similar to Mikosch and Patchkovskii \cite{Mikosch2013b}) and the blue lines including error bands ($\pm\sigma$) are the reconstructed spectra. {The parameters are $\xi_e=\xi_i=0.5$, {$\lambda$=1.5}, and ${\cal N}_{p}=10^6$. The blue lines in (a) and (b) show a parent and a fragment spectrum obtained by the Bayesian approach but without taking false coincidences into account. In (c) and (d) the false coincidences are treated in the correct way and the spectra can be reconstructed. The reconstruction works even better if the number of data points and therefore ${\cal N}_{p}=10^7$; see spectra (e) and (f).}}
    \label{fig1_mock}
\end{figure*}

In this section we demonstrate the performance of our algorithm.
It is  recalled that the reconstruction of $q^{(2)}$ is hampered by two disturbing influences: false coincidences and \new{pump-only background}.
The false coincidences are due to the presence of fragment molecules 
and imperfect detectors, so that the detected electron-ion pair does not necessarily belong to the same molecule.
To test the reconstruction power of our approach, we will treat these influences separately.
First, we study the case of false coincidences without  background signal $q^{(1)}$, i.e., we use  $\lambda_1\rightarrow 0$. The same problem has been addressed by Mikosch and Patchkovskii \cite{Mikosch2013b}.
They suggest to use a steplike spectral function for the parent and a series of Gaussian peaks for the spectral function of the fragment. These spectra are depicted as solid lines in  Fig. \ref{fig1_mock}. The left (right) column belongs to the parent (fragment) spectrum.
On purpose, the problem is aggravated by exponentially distributed step heights to  
study the impact  of false coincidences on the reconstructed spectrum if the parent-to-fragment ratio varies over several orders of magnitude. For comparison with the results of Mikosch and Patchkovskii, we use the same test spectra and the same parameters, namely   
$\xi_e=\xi_i=0.5$, and {$\lambda$=1.5}.
In general, the probability distribution for the energies $E_{\nu}$ corresponding to ion mass $M_{\mu}$ is given by the conditional probability
\begin{equation}\label{eq:cond:prop}
 q_{\nu|\mu} = \frac{q_{\mu\nu}}{q_{\mu\cdot}}\;.
\end{equation}
In the present test case we have two different ion masses (parent and fragment), for which we assume {according to Ref. \cite{Mikosch2013b} the probabilities $q_{\mu=1\cdot}=0.986$ and $q_{\mu=2\cdot}=0.014$ for the two species. }
\new{The spectrum of the parent (fragment) molecules is $q_{\text{p}} = q_{\mu=1 \nu}$ ($q_{\text{f}}=q_{\mu=2 \nu}$).} The product rule [inversion of \eq{eq:cond:prop}] yields
\begin{equation}
 q_{\mu\nu} = q_{\nu|\mu}\;q_{\mu\cdot}\;.
\end{equation} 

The mock data are generated as follows. In total we generate ${\cal N}_{p}$ measurements. For each measurement a random number $m$ is drawn from a Poisson distribution with mean $\lambda=1.5$. 
Next, $m$ index pairs $(\mu\nu)$ are generated according to the probability $q_{\mu\nu}$ .
Then, with probability $\xi_{i}$ the ion with mass $M_{\mu}$ is ``detected'' and added to the list of detected masses. Likewise, 
with probability $\xi_{e}$ the electron with energy $E_{\nu}$ is ``detected'' and added to the list of detected electron-energies. 
From this list we obtain the datasets $D_{1}$ and $D_{2}$ on which our approach is based. 
The results of the Bayesian reconstruction  are shown in 
Fig. \ref{fig1_mock}.
In the first row [Fig. \ref{fig1_mock}(a)] we used our algorithm ignoring the presence of false coincidences {by setting $\kappa = 0$.}
Therefore, the result is similar to that obtained by Mikosch and Patchkovskii \cite{Mikosch2013b}. The parent spectrum shows peaklike features from the fragment spectrum at the lower plateaus. Also the fragment spectrum includes contributions from the parent spectrum and has therefore a higher magnitude for the gaussian peak at the highest plateau in the parent spectrum.

Taking the false coincidences properly into account yields the results depicted in the second row [Fig. \ref{fig1_mock}(b)]. We see that the approach is able
to reassign the false coincidences to the spectrum they belong to.
Only at the lowest plateau, the error bars in the parent spectrum are comparable to the signal size, which just indicates that there are not enough data points. Increasing the number of data points produces result in Fig. \ref{fig1_mock}(c). Now  also the lowest plateau in the parent spectrum is reconstructed satisfactorily. 

\begin{figure*}
    \centering
    \includegraphics[width=1.9\columnwidth,angle=0]{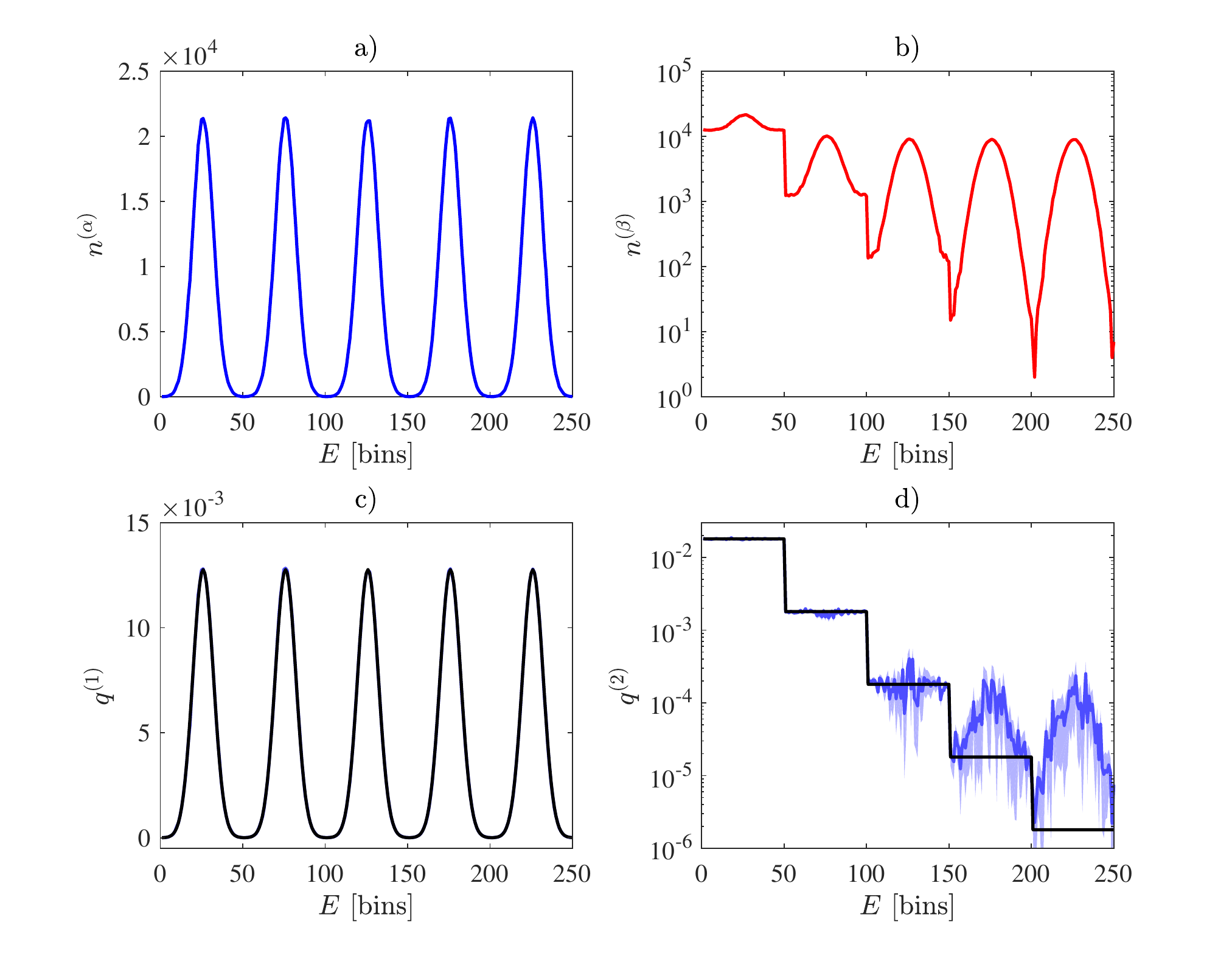}
    \caption{Simulated test spectra. {The subplots (a) and (b) represent the mock data for the pump-only (background) and the pump-probe (background + signal) measurement, respectively. The reconstructed background and signal are shown in (c) and (d). The black solid lines show the underlying test-spectra.} The blue jagged line along with the shaded regions represent the reconstructed signal and the error band ($\pm\sigma$). The parameters are $\lambda_1= \lambda_2=1.5$ and $\xi_i=\xi_e=0.5$.}
    \label{fig2_mock}
\end{figure*}

We see that the Bayesian approach is well suited to reassign false coincidences.
Next we will test, how the approach can handle the background subtraction. 
To this end we use the fragment spectrum from above (Gaussian peaks) as background spectrum and the parent spectrum from above (steps) as signal spectrum. This allows us to easily to identify residual  background structure in the reconstructed spectrum.
We choose the parameters $\lambda_1= \lambda_2=1.5$, $\xi_i=\xi_e=0.5$, and ${\cal N}_{p} = {\cal N}^{(\alpha)}_{p} = {\cal N}^{(\beta)}_{p}=10^7$. First we analyze the \new{PDF} 
$p(\pi\CS D_{2},\BK)$ of \eq{eq:parameters} for the parameters. The mean values and the $95~$\% confidence intervals are shown in Table \ref{tab:my_label_para}.
\begin{table}[]
    \centering
    \begin{tabular}{l||c|c}
      & $\pi$ & $\hat \pi$ \\
    \hline \hline
     $\lambda_1$ & $1.5$& $1.4996  \pm  0.0014$ \\
     $\lambda_2$ & $1.5$ & $1.5018  \pm  0.0020$ \\
     $\xi_i$ & $0.5$ & $0.4998  \pm  0.0004$ \\
     $\xi_e$ & $0.5$ & $0.4997  \pm  0.0004$ \\
    \end{tabular}
    \caption{Estimated parameters $\hat \pi$ with $95~$\% confidence intervals corresponding to the reconstructed spectra in Fig. \ref{fig2_mock}. The desired values $\pi$ are all within the confidence intervals.}
    \label{tab:my_label_para}
\end{table}
Obviously, all parameters are well estimated by the algorithm since the desired values are within the $95~$\% confidence intervals of the parameter's distributions.
The results of the reconstruction of the spectrum is depicted in Fig. \ref{fig2_mock}.
The simulated data $n^{(\alpha)}$ and $n^{(\beta)}$ are given in the upper part, and  the \new{``true''} and the reconstructed background and signal spectra in the lower part of the figure.
The background spectrum has a maximal standard error of $\sigma_{\max} \approx 10^{-4}$ at the tops of the background peaks. Therefore, it is reasonable that the signal spectrum in the center of the steps can only be reconstructed reliably if  it is larger than  this noise level of $10^{-4}$. We want to comment that the noise level can always be reduced by increasing the number of measurements ${\cal N}_{p}$. Consequently, the reconstruction works very well in the first three plateaus.
There the highest signal-to-background ratio of $2.4~\%$ is at $E = 125$.
Even at plateau four, where the signal is  merely $1.8\times 10^{-5}$, the reconstruction is satisfactory within the error band. At the last plateau, however, which merely has a size of $1.8\times 10^{-6}$, the center of the step cannot be reliably reconstructed.
At this point, a few comments are in order: Intuitively, the reader might be slightly disappointed about the  fact that  there is still background structure in the reconstructed signal in the last two steps. First of all, one has to be aware of the logarithmic scale and the fact that we are talking about the reconstruction of a signal that is only $0.02\%$ of the background. What we see in Fig. 
\ref{fig2_mock} is the best and most unbiased {\it form-free} reconstruction based on the data and the back ground information possible. 
We might nevertheless be disappointed, because we know that the true signal is flat and the structure we see has the form of the background. This disappointment is based on additional prior knowledge about signal and background structure, that we have withheld the Bayesian analysis.
Generally speaking, as long as we are seriously disappointed about a reconstruction, we have missed to incorporate parts of the background information.
A form-free reconstruction is only reasonable if we have no prior knowledge about the spectrum, whatsoever. Even if we only know that the spectrum has to be smooth, and should not jump discontinuously between neighboring bins, this represents prior knowledge that can be included, e.g., by a spline-based reconstruction or derivative priors \cite{linden_bayesian_2014}.

\section{Application to experimental data}
\label{sec:application}

  In this section we apply the Bayesian formalism, which we have derived and  tested for reliability in the previous sections, to pump-probe photoionization spectra of isolated acetone molecules. In recent studies of nonadiabatic relaxation processes triggered by photoexcitation of high-lying molecular Rydberg states, we were able to observe the time-dependent population transfer through internal conversion from the photoexcited states to a series of lower Rydberg states, revealing insight into the coupling of excited molecular states~\cite{Koch2017a,Maierhofer2016} and the corresponding fragmentation behavior~\cite{Koch2017}. 
The analysis of the pump-probe PEPICO spectra, however, is difficult in this case because of the following reasons: (i) The applied three-photon excitation scheme results in a strong pump-only background, and (ii) the pump-probe signal cannot be spectrally separated from the pump-only signal.
The application of a first and simplified version of the Bayes algorithm allowed us to assign photoelectron bands and to model the population transfer~\cite{Koch2017a}, which was limited by uncertainties about false coincidences. To demonstrate the superiority of the presented Bayesian approach, we show in the following PEPICO measurements on acetone molecules for selected pump-probe time delays. The Bayesian analysis of the spectra provides information about the relaxation and fragmentation dynamics that are enabled by the significant increase of the signal-to-noise ratio and the exclusion of false coincidences.

Table \ref{tab:expdata} lists the mean numbers of ionization events, $\lambda_1$ and $\lambda_2$, as well as the detection probabilities  $\xi_i$ and $\xi_e$ for selected pump-probe time delays of 12.5, 300, and 900 fs, as estimated from the corresponding data sets $D_2$. 
\new{Minor variations of the estimated parameter $\lambda_1$ result, due to statistical correlations in the parameter estimation, in variations of}
$\xi_i$ and $\xi_e$, although these values should remain constant. The decrease of $\lambda_2$ represents the decay of Rydberg state population~\cite{Koch2017a}.

\begin{table}[h!]
    \centering
    \begin{tabular}{l||c|c|c}
      & $\hat\pi$ ($12.5~$fs) & $\hat\pi$ ($300~$fs)  & $\hat\pi$ ($900~$fs) \\
    \hline \hline
     $\lambda_1$ & $0.3214  \pm  0.0021$ & $0.3562 \pm 0.0024$ & $0.3387 \pm 0.0023$\\
     $\lambda_2$ & $0.6446  \pm  0.0046$ &$0.1162  \pm  0.0024$ & $0.0679  \pm 0.0020$ \\
     $\xi_i$ & $0.2483  \pm  0.0015$ & $0.2215 \pm  0.0016$ & $0.2265 \pm 0.0017$  \\
     $\xi_e$ & $0.3735  \pm  0.0021$ & $0.3677 \pm   0.0024$ & $0.3647   \pm 0.0024$ \\
    \end{tabular}
    \caption{Estimated parameters $\lambda_1$, $\lambda_2$, $\xi_i$, and $\xi_e$ for the acetone measurements at different pump-probe delay times with $95~$\% confidence intervals.}
    \label{tab:expdata}
\end{table}

\new{Furthermore, we tested by mock data analysis that the fluctuations of $\lambda_1$ and $\lambda_2$ due to instabilities of the laser in our experiments have no influence on the conclusions drawn from the experimental data.}

\begin{figure*}
    \centering
    \includegraphics[width=1.9\columnwidth,angle=0]{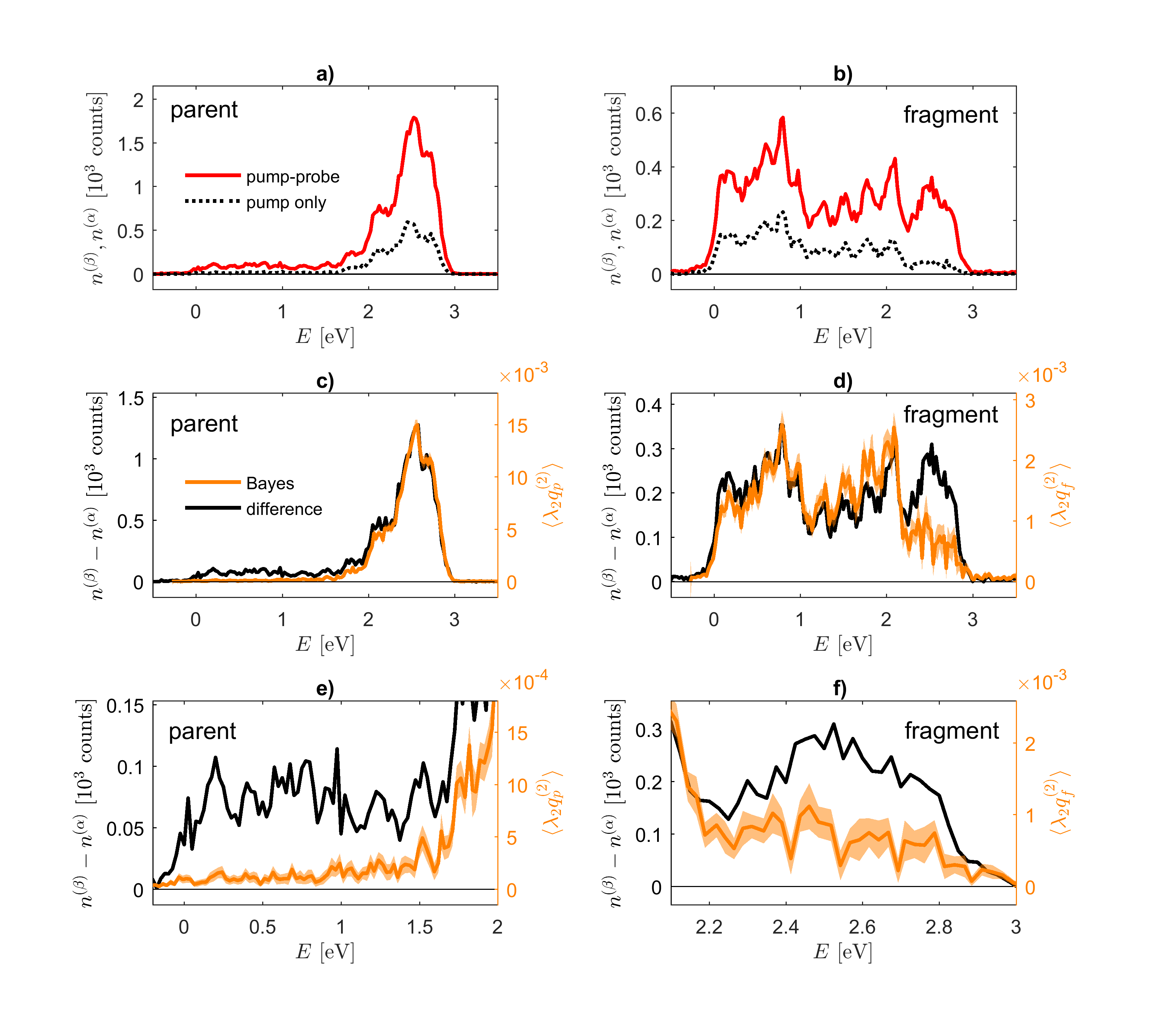}
    \caption{Photoelectron-photoion coincidence spectra obtained by pump-probe multiphoton ionization of acetone molecules with a pump-probe delay time of 12.5 fs.
    {The $x$ coordinates denote the measured electron kinetic energy $E$.}
    Graphs \new{(a) and (b)} show pump-only (black dashed lines) and pump-probe (red solid) spectra measured in coincidence with the parent and fragment, respectively. Graphs \new{(c) and (d)} depict the reconstructed spectra obtained with the Bayesian formalism (orange solid), together with the difference spectra obtained by subtraction of the pump-only from the pump-probe experiments (black solid), also for the parent (c) and fragment (d).
    \new{The shaded orange area indicates the error band ($\pm\sigma$), as obtained from the Bayes method.}
    Graphs \new{(e) and (f)} show {selected energy regions of (c) and (d), where the deviation of the difference spectra and the Bayesian spectra is significant, for the scaling of the $y$ axes of the two spectra as described in the text.}}
    \label{fig:spec12p5}
\end{figure*}

\begin{figure*}
    \centering
    \includegraphics[width=1.9\columnwidth,angle=0]{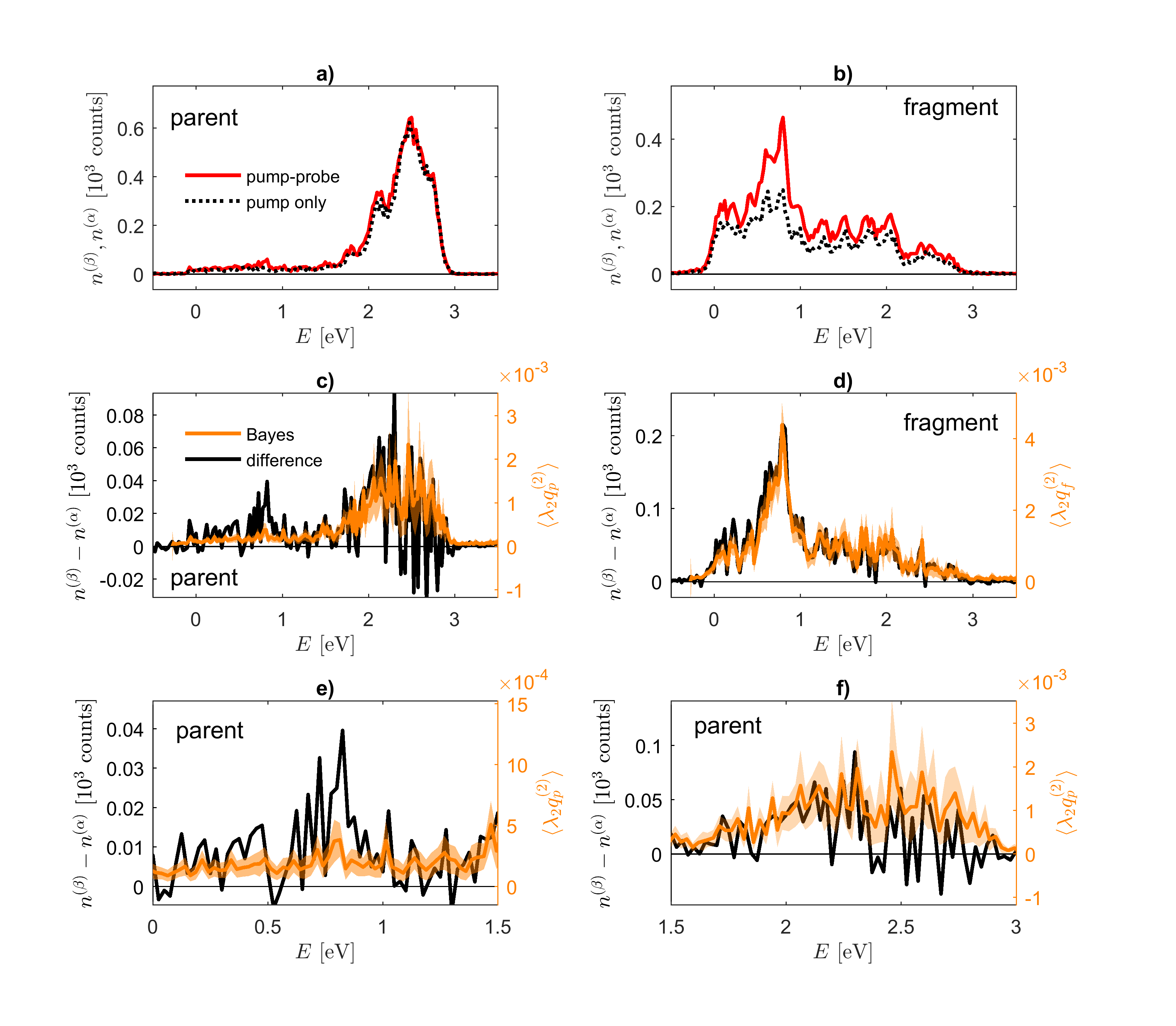}
    \caption{Same as described in the caption of Fig. \ref{fig:spec12p5} for a pump-probe delay time of $300~$fs.}
    \label{fig:spec300}
\end{figure*}

\begin{figure*}
    \centering
    \includegraphics[width=1.9\columnwidth,angle=0]{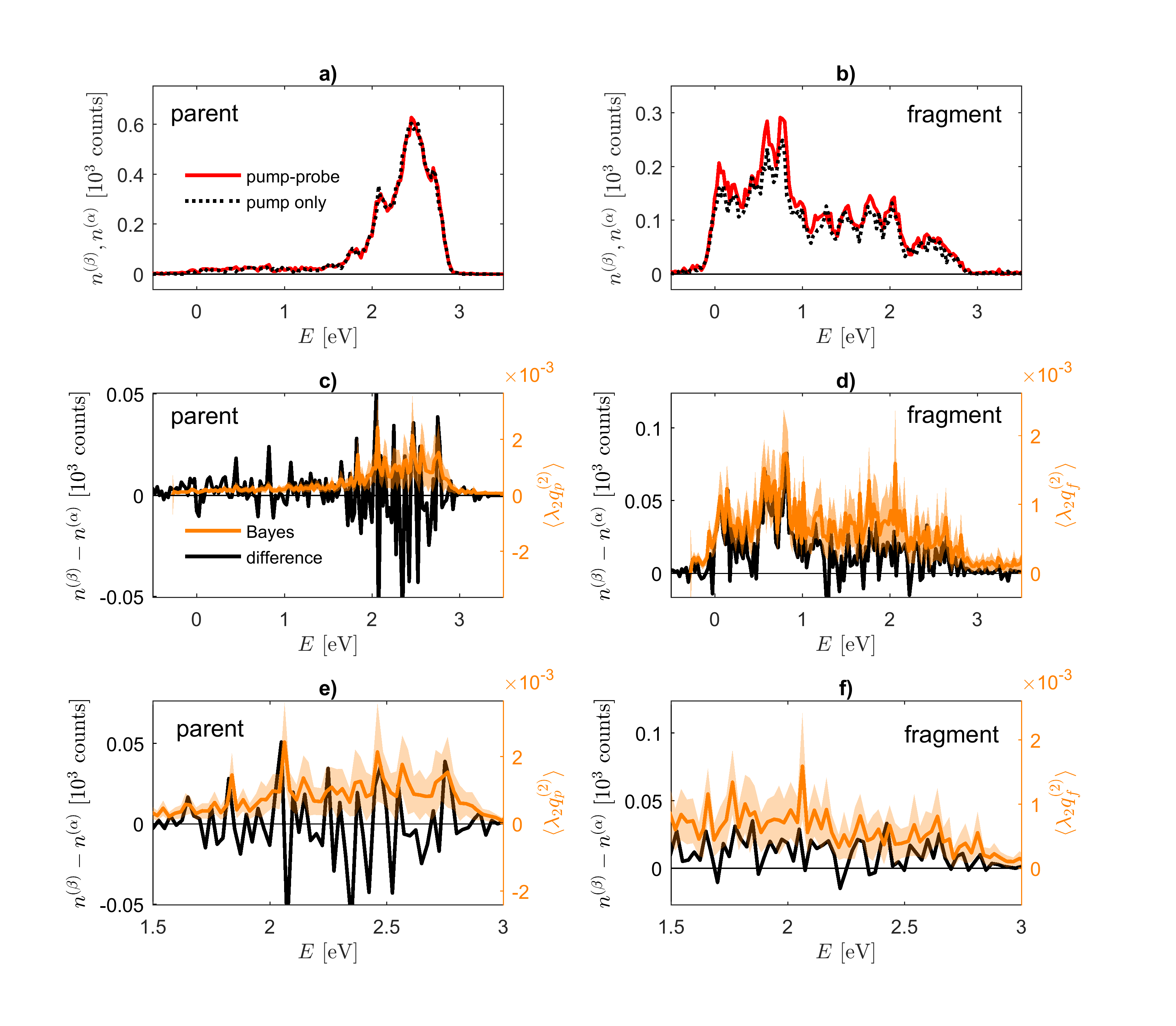}
    \caption{Same as described in the caption of Fig. \ref{fig:spec12p5} for a pump-probe delay time of $900~$fs.}
    \label{fig:spec900}
\end{figure*}

Figures \ref{fig:spec12p5}-\ref{fig:spec900} show time-resolved PEPICO measurements for the different time delays. Each of the three figures consists of six graphs showing photoelectron (PE) spectra detected in coincidence with parent (acetone) and fragment (acetyl) cations, as obtained in pump-probe and pump-only measurements (panels a and b), as well as a comparison of spectra that were reconstructed by the Bayesian algorithm to difference spectra obtained by a simple subtraction of pump-only and pump-probe measurements (panels c-f).
In the following we briefly review the assignment of PE bands in the spectra and previous interpretations of the relaxation and fragmentation dynamics, as discussed in more detail in Refs.~\cite{Koch2017a,Koch2017,Maierhofer2016}. Photoexcitation to high-lying Rydberg states (6$p$, 6$d$, 7$s$; PE peak at 2.7 eV) results in fast (320 fs) relaxation of the photoexcited population to lower Rydberg states and even faster population decay (80--130 fs) out of these states. These non-adiabatic internal conversion processes are mediated by Rydberg-valence couplings. The accompanying conversion of electronic energy to vibrational energy was found to cause fragmentation in the ionic state, that is after ionization, to acetyl ions and neutral methyl radicals, if the amount of converted energy exceeds $(0.79\pm0.04)$~eV. These relaxation dynamics give rise to the following PEPICO structures: The dominant parent PE band between 2 and 3 eV (panels a) results from photoionization of the photoexcited states and higher Rydberg states that are populated by internal conversion but for which the activation energy for fragmentation has not been reached. The fragment spectra (panels b), by contrast, consist of several PE bands up to ~2 eV, representing the Rydberg manifold down to the 3$p$ states, for which sufficient energy is converted for fragmentation. 
As information about the excited-state dynamics is contained in the signal associated with channel 2 (c.f., Fig. \ref{fig:scheme}), we now turn to the graphs in panels c-f of Figs. \ref{fig:spec12p5}-\ref{fig:spec900}, where the Bayesian results (orange lines) are compared to difference spectra (pump-probe minus pump-only, black).
\new{Note that the difference spectra are obtained in absolute counts, as displayed on the left ordinate, while for the reconstructed Bayesian spectra the expectation value $\lambda q_{\mu\nu}$ is plotted on the right ordinate. 
{ The reason for plotting these quantities is that the counts $n_{\mu\nu}$ are an adequate estimator for $\lambda q_{\mu\nu}$ under certain conditions. For instance, in the case of perfect detectors ($\xi_e=\xi_i=1$) we can set $n_{\mu\nu}$ to be proportional to the probability of the true coincidences in equation (48) and for small $\lambda$ the equation $n_{\mu\nu}\propto \lambda  q_{\mu\nu}$ follows.}
Since there is no consistent way to relate these quantities, we scale the {$y$-axis of the two spectra} to each other, {guided by our eyes,}  such that we obtain apparent overlap.}
The significant increase of the signal-to-noise ratio and the exclusion of false coincidences allows some interpretations, which were previously impossible and will be discussed in the following.

For the shortest pump-probe delay of 12.5 fs (Fig. \ref{fig:spec12p5}) strong signals are observed because the pump and probe pulses overlap in time and the higher light intensity results in a strong increase of the highly non-linear ionization process. At this delay, the agreement of the Bayesian spectra with the difference spectra is fair. Although the Bayesian and difference spectra cannot be compared quantitatively, scaling the spectra as shown in Figs. \ref{fig:spec12p5}(c) and \ref{fig:spec12p5}(d) indicates significant deviations, for the parent spectrum predominantly below 1.5 eV [Fig. \ref{fig:spec12p5}(e)] and for the fragment spectrum above 2.2 eV [Fig. \ref{fig:spec12p5}(f)].
We attribute the differences to false coincidences, which are \new{not included} in the Bayesian spectrum. While this effect is particularly pronounced at the shortest delays due to the strong signal, corresponding to high $\lambda$ values, it can also be observed in the 300 fs delay parent spectrum at ~0.8 eV [corresponding to the 3$p$ states, Fig. \ref{fig:spec300}(e)]. A high accuracy of the parent signal is a prerequisite for the determination of the fragmentation ratio, which is of importance in photofragmentation studies and could not previously be determined in the mentioned spectral regions because the influence of false coincidences was not clear. For the 12.5 fs delay measurement at 0.8 eV electron energy the deviation is most significant: Based on the Bayesian analysis the fragmentation probability is $(91\pm2)$\%, while the difference spectrum suggests a much lower value of 74~\% (both values are obtained by integrating the fragment and parent spectra between 0 and 1.5 eV).
\new{We note that these quantitative conclusions are obtained by consistently relating parent and fragment signals obtained within the same method, that is difference-parent to difference-fragment and Bayes-parent to Bayes-fragment. Consequently, it is not possible to quantitatively compare, for example, the Bayes-fragment spectrum to the difference-fragment spectrum, whereas the Bayesian fragment-to-parent ratio can well be compared to that of the difference method.}
The reliable Bayesian result shows that the nonadiabatic relaxation process from the photoexcited Rydberg states (6$p$, 6$d$, 7$s$) to the 3$p$ Rydberg states leads to almost complete fragmentation. The significant deviation demonstrates that a correction for false coincidences is required to obtain a reliable fragmentation probability, in particular, at these high $\lambda$ values. 
Similarly, although less pronounced, in the 300 fs delay measurement (Fig. \ref{fig:spec300}) \new{the Bayesian analysis yields a
fragmentation probability for the 3$p$ state of $(92\pm2)$~\%, which in this case even agrees with the value of 90~\%
obtained by signal subtraction.}

A similar deviation, although connected to a different interpretation, is  obvious in the 12.5 fs measurement in the fragment signal between 2.2 and 3.0 eV [Fig. \ref{fig:spec12p5}(f)].
{ Although the fragment signals obtained by the Bayesian formalism (orange) and by the difference approach (black) overlap up to 2.2~eV [Fig. \ref{fig:spec12p5}(d)], they deviate above 2.2~eV [Fig. \ref{fig:spec12p5}(f)]. The corresponding parent signals (Bayes and difference), in contrast, overlap between 2.2 and 3.0 eV [Fig. \ref{fig:spec12p5}(c)]. This indicates that the true probability for fragmentation, as obtained from the Bayesian analysis, is actually lower as suggested by the difference spectra, leading to the following interpretation: }
The PE band between 2.2 and 3.0 eV corresponds to ionization of Rydberg states that are populated directly by photoexcitation or by nonadiabatic relaxation, although with too little energy conversion for fragmentation~\cite{Koch2017}. 
Nevertheless, a certain fragment signal is present in this energy range, although at significantly different intensity, as predicted by Bayesian analysis compared to simple signal subtraction. This difference is important because the fragment signal can be caused by a subsequent fragmentation channel~\cite{Sandor2014} and the fragment signal strength is consequently a measure for the contribution of this fragmentation channel. In subsequent fragmentation the molecule is photoionized to the cationic ground state, which would not lead to fragmentation, but subsequently absorbs a further photon in the ionic state, which deposits sufficient energy for fragmentation~\cite{Sandor2014}. 
According to the Bayesian spectrum [Fig. \ref{fig:spec12p5}(f)], the fragmentation probability is $(6.7\pm1.6)$\%, compared to 20\% as obtained by the difference spectrum. Again, the incorrectly high fragment difference signal can be attributed to false coincidences, in this case of a fragment ion and an electron that belongs to a parent ion. We note that, first, although this subsequential pathway was identified in a previous experiment~\cite{Koch2017}, the corresponding branching ratio could not be determined. Second, the branching ratio sensitively depends on the laser intensity and pulse duration, as it is proportional to the probability of photon excitation in the cationic state. 

Next, we turn to another problem that is encountered when the pump-only spectrum is simply subtracted from the pump-probe spectrum. Because the pump-only measurement ($\alpha$, c.f. Fig. \ref{fig:scheme}) overestimates the pump-only signal contribution (channel 1) of a pump-probe measurement ($\beta$), the difference spectrum can become smaller than the actual signal from the excited state. This effect can be clearly seen in Figs. \ref{fig:spec900}(c) and \ref{fig:spec900}(e). At these long delay times the excited-state population has vastly decayed, resulting in a weak remaining parent signal, as correctly reconstructed by the Bayesian algorithm. The difference spectrum, by contrast, shows a significantly poorer signal-to-noise ratio and averages approximately to zero.

Finally, the poor signal-to-noise ratio of the difference spectra at the longest time delays prevents the identification of persistent signals, that can clearly be identified in the Bayesian analysis. The signals in the 900 fs (Fig. \ref{fig:spec900}) and 1500 fs (not shown) parent spectra between 1.5 and 3 eV [Fig. \ref{fig:spec900}(e)] and the fragment signal between 0 and 3 eV [Figs. \ref{fig:spec900}(d) and \ref{fig:spec900}(f)] are both clear indicators for a fraction of the population that does not decay from the excited Rydberg states. These nondecaying signals are consistent with previous two-photon excitation experiments~\cite{Maierhofer2016}, and could now also be identified in three-photon excitation experiments.

\section{Summary and Outlook}

We have demonstrated how Bayesian probability theory can be used  to analyze  pump-probe photoionization experiments with photoelectron-photoion coincidence detection. \new{The intrinsic problems of pump-only and/or probe-only background signals and false
coincidences originating from multiple ionization events can consistently be overcome.}
Most importantly, data acquisition times can be reduced significantly by the Bayesian analysis, as the correction for false coincidences allows for much higher ionization rates, and because it provides higher signal-to-noise ratios.

Based on challenging mock data we have demonstrated and quantified the reliability of the Bayesian method for spectral reconstruction. The application of the method to time-resolved PEPICO studies provided insights into the non-adiabatic relaxation dynamics of isolated acetone molecules.  Quantitative statements about fragmentation probabilities only became possible because false coincidences are taken into account correctly. The signal-to-noise ratio improvement was demonstrated by comparison to simple difference spectra and allowed us to identify ionization signals with a constant character, which could not be identified before in these data. Finally, also the problem of overestimating pump-only contributions in difference spectra, which is avoided in the Bayesian approach, could be demonstrated.

The Bayesian approach is highly flexible and is not at all restricted to the assumption made in the present paper.  It is straight forward to adjust it to different experimental conditions, such as 
fluctuating laser intensities, or to 
incorporate additional assumptions about the desired spectrum, such as smoothness, or 
more complicated fragmentation and excitation channels. 

In view of time-resolved photoionization experiments, the application of probe photon energies of about 15 to 20 eV is appealing because it exceeds the ground state ionization potential of most molecules and therefore allows to observe and follow the decaying photoexcited population all the way to the ground state. Femtosecond laser pulses in this energy range can be obtained from high-order harmonic generation~\cite{Koch2014c}. The related high probe-only background can be superimposed on the excited-state signal, leading to significant spectral distortions~\cite{Koch2015,Koch2014c}. The presented Bayesian approach can be adapted to such experimental conditions and will lead to similar improvements as presented in the present work.

\section{Acknowledgments}

The authors thank P. Maierhofer and M. Bainschab for fruitful discussions in the early stages of this project.
This work was partially supported by the Austrian Science Fund (FWF) under Grant No. P29369-N36, as well as by
NAWI Graz. 

\section{Appendix}

\subsection{Probability for a single coincidence \label{app:a}} 

Here we compute the probability $P(\text{SC}, E_{\nu},M_{\mu}\CS q_{\mu\nu},\lambda,\BK)$ that a single coincidence ($\text{SC}$) is detected in a single measurement, where the electron has energy $E_{\nu}$ and the ion has mass $M_{\mu}$. From the experimental observation \new{we cannot tell true from false coincidences}. As a first step we introduce the number $m$ of elementary events that occur in one measurement via the marginalization rule
\begin{align}
P(\text{SC}, E_{\nu},M_{\mu}\CS q ,\lambda,\BK) = 
\sum_{m=1}^{\infty}&P(\text{SC}, E_{\nu},M_{\mu}\CS m, q ,\lambda,\BK) \; \notag \\
&\times P(m\CS \lambda,\BK)\;.
\end{align}
The sum begins with $m=1$, otherwise there cannot be a coincidence.
There are two possibilities: (a) it is a true coincidence $\text{SC}_{t}$ or (b) it is a false one $\text{SC}_{f}$.
In the first case, the proposition says:
One of the events has $(\mu\nu)$, and both electron and ion are detected. 
At the same time there are
$m-1$ events with arbitrary $(\mu'\nu')$ for which the electrons and ions are not detected. 
The probability that the generated electron is detected is $\xi_{e}$ and the complement
$\oxi_{e}$ is the probability that it is  not detected. The analog quantities for the ions are $\xi_{i}$ and $\oxi_{i}$.
Taking into account that there are $m$ possibilities where the detected event may happen, we find
\begin{widetext}
\begin{align}
 P(\text{SC}_{t}, E_{\nu},M_{\mu}\CS m,q ,\lambda,\BK) &=
m\; \xi_{e}\xi_{i} q _{\mu\nu} \big(\overline\xi_{e}\overline\xi_{i}\big)^{m-1} \;, \\
 P(\text{SC}_{t}, E_{\nu},M_{\mu}\CS q ,\lambda,\BK) &=
q_{\mu\nu} \; \xi_{e}\xi_{i} \;\sum_{m=1}^{\infty} m \big(\overline\xi_{e}\overline\xi_{i}\big)^{m-1}
\frac{\lambda^{m}}{m!}e^{-\lambda}
= \lambda  \; \xi_{e}\xi_{i}\;q_{\mu\nu}\; e^{-\lambda\big( 1- \overline\xi_{e}\overline\xi_{i}\big)}\;. \nonumber
\end{align}
\end{widetext}
In the other case of false coincidences, the detected particles stem from two different events,
one with $\mu\nu'$ which yields the mass and one with $\mu' \nu$ from which the electron is detected.
The partner index can have any value, since it is not detected. In this case there are $m$ possibilities for the
position of the first event and $m-1$ for the second. \new{After introducing the marginal probabilities
\begin{equation}
q_{\cdot\nu} := \sum_{\mu'}q _{\mu'\nu}\;,\qquad
q_{\mu\cdot} := \sum_{\nu'}q _{\mu\nu'}\;.
\end{equation}
we have
\begin{widetext}
\begin{align}
P(\text{SC}_{f}, E_{\nu},M_{\mu}\CS m,q ,\lambda,\BK) &=\sum_{\mu' \nu'}\Theta (m\ge  2)\;
m(m-1)\; \xi_{e}\xi_{i} q _{\mu\nu'} q _{\mu'\nu} \big(\overline\xi_{e}\overline\xi_{i}\big)^{m-1} \nonumber \\
&=\Theta (m\ge 2)\;
m(m-1)\; \xi_{e}\xi_{i} q_{\mu\cdot} q_{\cdot\nu} \big(\overline\xi_{e}\overline\xi_{i}\big)^{m-1} \;,\\
P(\text{SC}_{f}, E_{\nu},M_{\mu}\CS q ,\lambda,\BK) &=
q_{\mu\cdot} q_{\cdot\nu}\; \xi_{e}\xi_{i} \;\sum_{m=2}^{\infty} m(m-1) \big(\overline\xi_{e}\overline\xi_{i}\big)^{m-1}
\frac{\lambda^{m}}{m!}e^{-\lambda} \notag \\
&= \lambda \; \xi_{e}\xi_{i}  q_{\mu\cdot}q_{\cdot\nu} \lambda \overline\xi_{e}\overline\xi_{i} e^{-\lambda\big( 1- \overline\xi_{e}\overline\xi_{i}\big)}\;,
\end{align}
\end{widetext}
where $\Theta$ is a generalization of the Heaviside step function for boolean arguments, as is used in some programming languages,
\begin{align}\label{eq:heaviside}
\Theta(b) &=
\begin{cases}
1&\text{if } b = \text{True}	\\
0&\text{if } b = \text{False}	\;.
\end{cases}
\end{align}}
In total we obtain
\begin{align}
P(\text{SC}, E_{\nu},M_{\mu}\CS q ,\lambda,\BK) = &\lambda \xi_{e}\xi_{i} \big( q_{\mu\nu} 
+ \lambda q_{\cdot\nu} q_{\mu\cdot} \overline \xi_{e}\overline \xi_{i}\big) \notag \\ \;&\times e^{-\lambda\big( 1-\overline\xi_{e}\overline\xi_{i} \big)}.
\end{align}
Marginalization over $\mu\nu$ yields the 
 probability for a single coincidence during a measurement, irrespective of  the measured electron energy and ion mass
\begin{align}
 q _{\text{SC}}:=P(\text{SC}\CS q ,\lambda,\BK) &=\sum_{\mu\nu} P(\text{SC}, E_{\nu},M_{\mu}\CS q ,\lambda,\BK) \notag \\
&=\lambda  \xi_{e}\xi_{I} \big( 1 + \lambda\overline \xi_{e}\overline \xi_{i} \big)\;
e^{-\lambda\big( 1-\overline\xi_{e}\overline\xi_{i}\big)}\;.
\end{align}
However, given that we only consider single coincidences, the probability for an outcome $\mu\nu$ in this coincidence is
\begin{align}
 \tilde q _{\mu\nu} &= P(E_{\nu},M_{\mu}\CS \text{SC},q ,\lambda,\BK) \notag \\
 &=
\frac{P(E_{\nu},M_{\mu},\text{SC}\CS q ,\lambda,\BK) }{P(\text{SC}\CS q ,\lambda,\BK) } \notag \\
 &=\frac{ q _{\mu\nu} 
+ \lambda \overline \xi_{e}\overline \xi_{i} q_{\cdot\nu} q_{\mu\cdot} }{1 +  \lambda \overline \xi_{e}\overline \xi_{i}}\;.
\end{align}
This is the required relation between the experimentally measured spectrum (distorted by false coincidences) $\{ \tilde q_{\mu\nu}\}$ and the underlying true spectrum $\{ q_{\mu\nu}\}$.

\subsection{Likelihood}
\label{app:c}

The sought likelihood function is multinomial if the total number of single coincidence events $N_{\text{SC}}$ is given:
\new{\begin{equation}
P(n\CS  \tilde q,N_{\text{SC}},\BK) =\Theta\big(\sum_{\mu\nu} n_{\mu\nu}=N_{\text{SC}}\big) N_{\text{SC}}! \prod_{\mu\nu} 
\frac{\big(\tilde q_{\mu\nu}\big)^{n_{\mu\nu}}}
{n_{\mu\nu}!}
\;,
\end{equation}
with the definition of $\Theta(.)$ given in \eq{eq:heaviside}.}
Since $N_{\text{SC}}$ is unknown, but the number of measurements ${\cal N}_{p}$ is given, we have to introduce $N_{\text{SC}}$ by the marginalization rule and obtain
\begin{widetext}
\begin{align}
 P(n\CS  \tilde q_{\mu\nu},{\cal N}_{p},\BK) &=
\sum_{N_{\text{SC}}} P( n \CS  \tilde q,N_{\text{SC}}) \;P(N_{\text{SC}}\CS {\cal N}_{p},q_{\text{SC}},\BK) \notag \\
&=\sum_{N_{\text{SC}}} \Theta\big(\sum_{\mu\nu} n_{\mu\nu}=N_{\text{SC}}\big)
\frac{N_{\text{SC}}!}{\prod_{\mu\nu} n_{\mu\nu}!} \prod_{\mu\nu} 
\big(\tilde q\big)_{\mu\nu}^{n_{\mu\nu}}\;
P(N_{\text{SC}}\CS {\cal N}_{p},q_{\text{SC}}) \notag \\
&=
\frac{N_{\text{SC}}^{*}!}{\prod_{\mu\nu} n_{\mu\nu}!} \prod_{\mu\nu} 
\big(\tilde q_{\mu\nu}\big)^{n_{\mu\nu}}\;
P(N_{\text{SC}}^{*}\CS {\cal N}_{p},q_{\text{SC}})\;,
\end{align} 
\end{widetext}
with $N^*_{\text{SC}}= \sum_{\mu\nu} n_{\mu\nu}$. We remark that the probability  that in ${\cal N}_{p}$ measurements $N_{sc}$ single coincidences are detected is the outcome of a Bernoulli experiment which is described by a binomial distribution,
\begin{equation}
P(N_{\text{SC}}\CS {\cal N}_{p},q_{\text{SC}}) = {\cal B}(N_{\text{SC}}\CS {\cal N}_{p},q_{\text{SC}})\;.
\end{equation}
Normalization leads to the sought likelihood.

\subsection{The Jacobian determinant}
\label{app:b}
Here, we will use \eq{eq:q:tilde:q}
to map the posterior $ p(\tilde q \CS n ,\oBK)$ of
\eq{eq:posterior:tilde:q} to the \new{PDF} $ p(q \CS n ,\oBK)$ in terms of the true spectrum $q$. The posterior 
\begin{equation}\label{eq:}
p(\tilde q \CS n ,\oBK) = \mathcal{M}(\tilde q)\;\delta\big( \tilde S-1 \big)
\end{equation}
consists of a multinomial part $\mathcal{M}(\tilde q)$ and a delta-function.
The desired transformation follows from
\begin{align}\label{app:aux1}
 p(q \CS n ,\oBK)  &= p(\tilde Q(q) \CS n ,\oBK) 
\det\bigg(\frac{d \tilde Q(q)}{dq}\bigg) \notag \\
&= \mathcal{M}(\tilde Q(q)) \delta\big(\tilde S-1  \big)  
\det\bigg(\frac{d \tilde Q(q)}{dq}\bigg)\;.
\end{align}
First we compute the Jacobian of the transformation
\new{\begin{align}
\frac{d \tilde Q(q)}{dq}&=
\big( 1+\kappa \big)^{-1} \bigg( \delta_{\mu\mu'}\delta_{\nu\nu'} + \kappa \Delta^{\mu\mu'}_{\nu\nu'}\bigg)\\
\text{with}\qquad\Delta^{\mu\mu'}_{\nu\nu'} &= \delta_{\mu\mu'} q_{\cdot\nu} + \delta_{\nu\nu'} q_{\mu\cdot}\;. \notag
\end{align}}
Then the determinant reads
\begin{equation}
\det\bigg(\frac{d \tilde Q(q)}{dq}\bigg) = (1+\kappa)^{-{\cal N}_{\mu} {\cal N}_{\nu}} \det\big( \uu + \kappa \Delta \big) \;.
\end{equation}
One can easily prove that with $M^{\mu\mu'}_{\nu\nu'} = q_{\mu\cdot}q_{\cdot\nu}$
we find
\begin{align}
\Delta^{2} &= \Delta + 2 M\;,\nonumber\\
\Delta M     &= 2 M\;,\nonumber\\
\Delta^{n}&= \Delta + \bigg( 2^{n}-2 \bigg) M\;,\nonumber\\
\text{tr}\big( \Delta \big) &= {\cal N}_{\mu}+{\cal N}_{\nu}\;,\nonumber\\
\text{tr}\big( M \big) &=1\;.
\end{align}
We can use these relations to calculate the logarithm of the remaining determinant:
\begin{widetext}
\begin{align}
\ln\bigg( \det\bigg( \uu + \kappa\Delta \bigg) \bigg) &= 
\text{tr}\bigg( \ln\bigg(\uu + \kappa \Delta  \bigg)\bigg) \notag\\
&=\sum_{n=1}^{\infty} \frac{(-1)^{n+1}}{n} \kappa^{n}
\text{tr} \big(\Delta^{n}\big) \notag\\
&=\sum_{n=1}^{\infty} \frac{(-1)^{n+1}}{n} 
\bigg[
\kappa^{n}\big({\cal N}_{\mu}+{\cal N}_{\nu}-2\big)+ (2\kappa)^{n}
\bigg] \notag\\
&=\big({\cal N}_{\mu}+{\cal N}_{\nu}-2\big) \ln\big( 1+\kappa \big)
+\ln\big( 1+2\kappa \big).
\end{align}
\end{widetext}
Hence,
\begin{equation}
\det\big( \uu +\kappa\Delta \big) = \big( 1+\kappa \big)^{{\cal N}_{\mu}+{\cal N}_{\nu}-1}   \frac{1+2\kappa}{1+\kappa} \;,
\end{equation}
and eventually we obtain
\begin{equation}\label{app:Jakobi}
\det\bigg(\frac{d \tilde Q(q)}{dq}\bigg)
 = \big( 1+\kappa \big)^{-({\cal N}_{\mu}-1)( {\cal N}_{\nu}-1)}
   \frac{1+2\kappa}{1+\kappa} \;.
\end{equation}

Finally we also want  to express the delta-function in \eq{app:aux1} in terms of the variables $q$. To this end, we express $\tilde S$ as function of \new{$S=\sum_{\mu\nu} q_{\mu\nu}$}
\begin{equation}\label{app:dummy}
\tilde S  =  \sum_{\mu\nu} \tilde q_{\mu\nu} 
= \sum_{\mu\nu} \frac{q_{\mu\nu}+\kappa q_{\mu\cdot}q_{\cdot\nu} }{1+\kappa} 
= \frac{S+\kappa S^{2}  }{1+\kappa}  \;.
\end{equation}
The argument of the delta-function $\delta(\tilde S-1)$ has a unique zero at $S=1$. Considered as function of $S$, we therefore have
\begin{equation}\label{app:delta}
\delta(\tilde S-1) = \frac{\delta(S-1)}{\big| \frac{d\tilde S}{dS}\big|} =
\frac{1+\kappa}{1+2\kappa}\;\delta(S-1)\;.
\end{equation}
Combination with \eq{app:Jakobi} and insertion in \eq{app:aux1} finally yields
\begin{equation}
p(q \CS n ,\oBK)  = \mathcal{M}(\tilde Q(q)) \;\delta\big(S-1\big)\;
\big( 1+\kappa \big)^{-({\cal N}_{\mu}-1)({\cal N}_{\nu}-1)}\;.
\end{equation}

\subsection{Probabilities for the count-pairs $(N_{e}, N_{i})$}
\label{app:Na} 
	
We consider the pump-only or the pump-probe experiment and ask for the probability 	
$P(N_i, N_e \CS \lambda, \xi_i, \xi_e)$ that during a single measurement $N_{e}$ electrons
and $N_{i}$ ions are detected, irrespective of their energy or mass, given the mean number $\lambda$ of elementary events during one pulse and given the detection probabilities $\xi_{e}$ and $\xi_{i}$. First we introduce the number $m$ of elementary events via the marginalization rule, exploiting the fact that detection of electrons and ions is uncorrelated, i.e.,  
\begin{align}
 P_{N_{i}N_{e}}&:=P(N_i, N_e \CS \lambda, \xi_i, \xi_e)\notag \\
 &= \sum_{m=0}^{\infty} 
 P(N_i \CS m,\xi_i, \BK) P(N_e \CS m,\xi_e, \BK) P(m\CS \lambda, \BK) \;.
 \label{eq:14}
\end{align}
The probability $ P(N_i \CS m,\xi_i, \BK)$ is binomial, 
since for each of the $m$ ions there is a probability $\xi_{i}$ that it will be detected.
The same holds true for the number of detected electrons, i.e.,
\begin{eqnarray}
 p(N_i \CS m,\xi_i, \BK) &= \mathcal{B}(N_i\CS \xi_i, m) \;,  \nonumber\\
 p(N_e \CS m,\xi_e, \BK) &= \mathcal{B}(N_e\CS \xi_e, m)\;.
\end{eqnarray}
The number of elementary events $m$ is Poisson distributed with mean $\lambda$,
\begin{equation}
 p(m \CS \lambda, \BK) =\mathcal{P}(m\CS \lambda )\;.
\end{equation}
The easiest way to compute the desired probabilities is via the generating function, which is defined as
\begin{widetext}
\begin{align}
 \Phi(x,y) &:= \sum_{N_{e}=0}^{\infty}\sum_{N_{i}=0}^{\infty}
x^{N_{e}} y^{N_{i}} P_{N_{e}N_{i}}  \nonumber\\
&= 
\sum_{m=0}^{\infty} e^{-\lambda} \frac{\lambda^{m}}{m!}  \;
\sum_{N_{e}=0}^{\infty} \; 
{m\choose N_{e}} (x \xi_{e})^{N_{e}} \overline \xi_{e}^{m-N_{e}}  
\sum_{N_{i}=0}^{\infty} \;
{m\choose N_{i}} (y \xi_{i})^{N_{i}} \overline \xi_{i}^{m-N_{i}}  \nonumber\\
&=\sum_{m=0}^{\infty} e^{-\lambda} \frac{\lambda^{m}}{m!}  \; \bigg[
\bigg( x \xi_{e}+\overline \xi_{e} \bigg)
\bigg( y \xi_{i}+\overline \xi_{i} \bigg)
\bigg]^{m} \nonumber\\
&=e^{-\lambda} \;e^{\lambda 
\big( x \xi_{e}+\overline \xi_{e} \big)
\big( y \xi_{i}+\overline \xi_{i} \big)
}.
\end{align}
\end{widetext}
The probabilities $P_{N_{e}N_{i}}$ are then readily obtained as coefficients of the Taylor expansion
\begin{equation}
P_{N_{e},N_{i}}
 = \frac{\big(\frac{\partial}{\partial x}\big)^{N_{e}} }{N_{e}!}\frac{\big(\frac{\partial}{\partial y}\big)^{N_{i}} }{N_{i}!}
\Phi(x,y)\bigg|_{x=0,y=0}\;.
\end{equation}
Straight forward evaluation of the derivatives, using, e.g., MATHEMATICA, yields for the lowest terms,
\begin{eqnarray}
P_{00} &= e^{-\lambda(1-\oxi_{e}\oxi_{i}) }  \nonumber\\
P_{10} &=\lambda \xi_{e}\oxi_{i}\; P_{00}  \nonumber\\
P_{01} &=\lambda \oxi_{e}\xi_{i}\; P_{00}  \nonumber\\
P_{11} &=\lambda \xi_{e}\xi_{i} \big(1 + \kappa\big)\; P_{00}  \nonumber\\
P_{20} &=\frac{\lambda^{2}}{2} \xi_{e}^{2}\oxi_{i}^{2}\; P_{00}  \nonumber\\
P_{02} &=\frac{\lambda^{2}}{2} \oxi_{e}^{2}\xi_{i}^{2}\; P_{00}  \nonumber\\
P_{30} &=\frac{\lambda^{3}}{3!} \xi_{e}^{3}\oxi_{i}^{3}\; P_{00}  \nonumber\\
P_{03} &=\frac{\lambda^{3}}{3!} \oxi_{e}^{3}\xi_{i}^{3}\; P_{00}  \nonumber\\
P_{21} &=\frac{\lambda^{2}}{2}  \xi_{e}^{2}\xi_{i}\oxi_{i} \; \big(2+\kappa\big)\;P_{00}  \nonumber\\
P_{12} &=\frac{\lambda^{2}}{2}  \xi_{e}\oxi_{e}\xi_{i}^{2} \; \big(2+\kappa\big)\;P_{00}\;,
\end{eqnarray}
with $\kappa=\lambda\oxi_{e}\oxi_{i}$.

\section*{References}

\bibliographystyle{apsrev4-1}
\bibliography{literatur}

\end{document}